\documentclass[12pt,a4paper]{article}

\usepackage{subeqn}
\usepackage{graphicx}
\setkeys{Gin}{width=0.9\textwidth,height=0.7\textheight,keepaspectratio}
\usepackage{hyperref}

\makeatletter
\def\@oddfoot{\hfill\footnotesize\sf Mei, Roberts \& Li, \today}
\makeatother

\newcommand{\bsub}{\begin{subequations}}
\newcommand{\esub}{\end{subequations}}

\newcommand{\F}{\mbox{\boldmath$ F$}}

\renewcommand{\u}{\mbox{\boldmath$ u$}}

\renewcommand{\vec}[1]{\mbox{\boldmath$#1$}}

\newcommand{\cE}{{\cal E}}
\newcommand{\cF}{{\cal F}}
\newcommand{\cG}{{\cal G}}

\newcommand{\cO}{{\cal O}}
\newcommand{\cM}{{\cal M}}

\newcommand{\pard}[2]{\frac{\partial {#1}}{\partial {#2}}}

\newcommand{\cL}{{\cal L}}
\newcommand{\bs}{\mbox{\boldmath$ s$}}

\newcommand{\cV}{{\cal V}}
\newcommand{\avu}{{\bar u}}
\newcommand{\avk}{{\bar k}}
\newcommand{\ave}{{\bar \epsilon}}
\newcommand{\Dt}[1]{\frac{\partial #1}{\partial t}}
\newcommand{\Dx}[1]{\frac{\partial #1}{\partial x}}

\newcommand{\cbrt}[1]{{#1^{1/3}}}
\newcommand{\cbrty}{{\textstyle \frac{4}{3}\cbrt{\left(\frac{y}{\eta}\right)}}}
\newcommand{\cubey}{{\textstyle \frac{4}{3}\cbrt{\left(\frac{y}{h}\right)}}}
\newcommand{\Ceo}{C_{\epsilon1}}
\newcommand{\Cet}{C_{\epsilon2}}
\newcommand{\se}{\sigma_\epsilon}
\newcommand{\sk}{\sigma_k}
\newcommand{\nut}{\tilde\nu}
\newcommand{\Cm}{C_\mu}
\newcommand{\Ord}[1]{{\cal O}\left(#1\right)}
\newcommand{\deltax}{{\vartheta}}
\newcommand{\keps}{$k$-$\epsilon$}
\newcommand{\lamb}{{\tilde\lambda}}
\newcommand{\vz}{\upsilon}
\newcommand{\ftt}{{2/3}}
\newcommand{\reduce}{{\sc reduce}}

\makeatletter
\newcounter{lineno}
\def\verbatimlisting#1{\setcounter{lineno}{0}%
    \begingroup{\footnotesize} \@verbatim \frenchspacing \@vobeyspaces
    \parindent=20pt
    \everypar{\stepcounter{lineno}\llap{\thelineno\ }}\footnotesize\input#1
    \endgroup
}
\makeatother

\begin{document}
\title{\sf Modelling the dynamics of turbulent floods}

\author{Z.~Mei\thanks{Department of Mathematics, University of
Marburg, 35032 Marburg/Lahn, Germany.
\protect\url{mailto:meizhen@mathematik.uni-marburg.de}}
\and A.J.~Roberts\thanks{Department of Mathematics \& Computing,
University of Southern Queensland, Toowoomba, Queensland 4350,
Australia.  \protect\url{mailto:aroberts@usq.edu.au}}
\and Zhenquan Li\thanks{Department of Mathematics \& Computing,
University of Southern Queensland, Toowoomba, Queensland 4350,
Australia.  \protect\url{mailto:zhen@usq.edu.au}}
}
\date{June 25, 1999}

\maketitle

\begin{abstract}
Consider the dynamics of turbulent flow in rivers, estuaries and
floods.
Based on the widely used \keps{} model for turbulence, we use the
techniques of centre manifold theory to derive dynamical models for
the evolution of the water depth and of vertically averaged flow
velocity and turbulent parameters.
This new model for the shallow water dynamics of turbulent flow:
resolves the vertical structure of the flow and the turbulence;
includes interaction between turbulence and long waves; and gives a
rational alternative to classical models for turbulent environmental
flows.
\end{abstract}


\paragraph*{AMS subjects:} 58F39, 76F20, 76B15.

\tableofcontents

\section{Introduction}
\label{sintro}

Consider the dynamics of a turbulent flow over ground, as occurs in
rivers, channels or floods.
In such flows it is the large scale horizontal variations which are
important; the vertical structure of velocity and turbulence may be
expected to be determined by the local conditions of the horizontal
flow.
In this situation we may seek a model of the flow which only involves
``coarse'' depth-averaged quantities.
Such models have been constructed before; for example, Fredsoe \&
Deigaard \cite[pp37--39]{Fredsoe92} depth-average the $k$-equation
model of turbulent flow to model the dynamics of breakers on a beach,
whereas depth-averaged {\keps} equations have been used by Rastogi \&
Rodi \cite{Rastogi78} to model heat and mass transfer in open
channels, and by Keller \& Rodi \cite{Keller88} to investigate flood
plain flows.
The need for such sophisticated models was also indicated by:
Peregrine \cite[p97]{Peregrine72} commenting that an empirical
friction law derived from channel flow underestimates the turbulence
in breakers and surf; and Mei \cite[p485]{Mei89} observing that eddy
viscosities need to be different in and outside of the surf zone.

However, the recent development of centre manifold theory and related
techniques presages a much deeper understanding of the process of
modelling nonlinear dynamics and foresees the systematic reduction of
many nonlinear problems to an underlying low-dimensional system.
For example, the process of depth-averaging has been shown to be
deficient as a modelling paradigm \cite{Roberts94c}.
In the context of turbulent flow, we show in \S\ref{slowd} how
the mean motions may be determined by a few critical modes which have
a non-trivial structure in the vertical, e.g.~as a first approximation
the horizontal velocity $u$ and turbulent energy density $k$ are taken
to have a cube-root dependence.
Moreover, the amplitude of these modes and their evolution may be
expressed in terms of depth-averaged quantities.
We derive the following coupled nonlinear set of equations to model
the turbulent, large-scale flow of water over ground,
see~(\ref{mdl3}): \bsub\label{mdli}
\begin{eqnarray}
\label{mdlih}
	\frac{\partial \eta}{\partial t}&\sim&
	- \frac{\partial(\eta\avu)}{\partial x}\,,
\\
\frac{\partial\avu}{\partial t}&\sim&
-1.030\frac{\nut\avu}{\eta^2}
+(0.0504-0.243\,\lamb)\frac{\nut\avu^3}{\eta^2\avk}
\nonumber\\  &&{}
+0.961\,g\left(\theta-\frac{\partial\eta}{\partial x}\right)
-1.105\,\avu \frac{\partial \avu}{\partial x}
\label{mdliu}
\,,\\
\frac{\partial\avk}{\partial t}&\sim&
-0.993\,\ave-0.0927\frac{\avk^3}{\eta^2\ave}
+(0.589+0.516\,\lamb)\frac{\nut\avu^2}{\eta^2}
-1.106\,\avu \frac{\partial \avk}{\partial x}
\label{mdlik}
\,,\\
\frac{\partial\ave}{\partial t}&\sim&
-2.101\frac{\ave^2}{\avk}
+(1.552-3.215\,\lamb)\frac{\nut\ave\avu^2}{\avk\eta^2}
\nonumber \\  &&{}
-0.173\,\lamb\ave\Dx\avu
+0.533\,\lamb\frac{\ave\avu}{\avk}\Dx\avk
-(1+0.735\,\lamb) \avu \frac{\partial \ave}{\partial x}
\label{mdlie}
\,.
\end{eqnarray}
\esub Here $\eta$ is the water depth and $\avu$, $\avk$ and $\ave$ are
depth averaged flow velocity, turbulent energy and turbulent
dissipation, respectively.
The other variables appearing are: $\nut=C_\mu\avk^2/\ave$ measuring
the local eddy diffusivity, see~(\ref{mdl3nu}); and
$\lamb=\eta^2\ave^2/\avk^3$ being proportional to the ratio of the
vertical mixing time to the turbulent eddy time, see~(\ref{Elamb}).
For example, in \S\S\ref{SSdam} this model is used to predict the flow
after a dam breaks.
See in Figure~\ref{fig:damhu} (p\pageref{fig:damhu}) the formation of
a turbulent bore rushing downstream from the dam.
The turbulence in the bore is generally highest near the front and
decays away behind as seen in Figure~\ref{fig:damturb}
(p\pageref{fig:damturb}).
This gives one example of the variations in spatial structure of the
turbulence that underlies shallow water flows.

Modelling turbulent flow is one of the major challenges in fluid
dynamics.
While large eddies, which can be as large as the flow domain, extract
energy from the mean flow and feed it into turbulent motion, the
eddies also act as a vortex stretching mechanism and transfer the
energy to the smallest scales where viscous dissipation takes place.
It is the scale at which the dissipation occurs which determines the
rate of energy dissipation.
However, the inflow of energy into turbulent motion is a
characteristic of only the large scale motion.
In other words, the turbulent but small scale motion is often
dominated and determined by the large scale motion and can be treated
as a perturbation of the mean flow.
The coupling of energy transportation and energy dissipation with the
mean flow is adequately described by widely used second-order closure
models.
In particular, the most popular choice is the two equation
$k$-$\epsilon$ model, see for example Launder {\em et al}
\cite{Launder75}, Hanjali\'c \& Launder~\cite{Hanjalic72},
Rodi~\cite{Rodi80}, and Speziale~\cite{Speziale91} for reviews.

We base our analysis upon the $k$-$\epsilon$ model (\S\ref{skemod})
for the turbulence underlying the free surface of fluid in a channel,
river or over a flood plain.
The resultant model~(\ref{mdli}) is basically a model for the
evolution of vertically averaged quantities; the model resolves
large-scale, compared to the depth, dynamical structures in the
horizontal.
It is important to note the difference between models obtained by
depth-averaged equations (which are known to be
incorrect~\cite{Roberts94c} for other similar long-wave dynamics), and
our model which is, for convenience, written in terms of
depth-averaged quantities.

In \S\ref{sss-homo} we derive the ``coarse'' low-dimensional
model~(\ref{mdli}) from the ``fine'' {\keps} equations.
Despite the well recognised limitations of the {\keps} equations as a
model of turbulence, we anticipate that the information retained in
our coarse model is reasonably insensitive to deficiencies in the
{\keps} dynamics.
Further modelling may be done via more sophisticated Reynolds stress
models for channel flow such as that described by Gibson \& Rodi
\cite{Gibson89}.
One aspect to note is that the model we derive has \emph{no}
adjustable parameters---all constants are determined from values
established for the {\keps} turbulence model and its boundary
conditions.
Thus the model predictions are definitive.
We describe some example solutions in \S\ref{seg} to illustrate
the dynamical predictions of the model.

Finally, in Appendix~\ref{Scom} we comment on the status of the theory
of centre manifolds in this development of a low-dimensional model of
turbulent flow using centre manifold techniques.

\section{The {\keps} model of turbulent flow}
\label{skemod}

Consider the two-dimensional inviscid $k$-$\epsilon$ model of
turbulent flow over rough ground.
Distance parallel to the ground's slope is measured by $x$, while we
denote $y$ as the direction normal to the slope.
Molecular dissipation is neglected because we anticipate little direct
effect for it in flood waves of a depth $h=\cO(\mbox{metre})$ over
ground with roughness which may be many times the length-scale of
viscous dissipation.
Turbulent eddies are proposed to be the dominant mechanism for
dispersion and dissipation.
We denote the ensemble mean velocity components and pressure by $u$,
$v$ and $p$ respectively, that is, for simplicity, we omit any
distinguishing overbars (instead overbars will later be used to denote
depth-averaged quantities).
Then the incompressible $k$-$\epsilon$ model (with ensemble means) may
be expressed as
\bsub\label{EAKeFW}
\begin{equation}
\left[\begin{array}{c}0\\\pard{\u}{t}\end{array}\right]=\left[
\begin{array}{c}\pard{u}{x}+\pard{v}{y}\\ \F(p, \u)
\end{array}\right],
\end{equation}
where the vector $\u=(u,v, \eta, k,\epsilon)$ \footnote{We adopt the
notation that a vector in parentheses, such as $(u,v, \eta,
k,\epsilon)$, is a short-hand for the corresponding column vector.} is
formed from the velocities $u$, $v$ in the horizontal and vertical
directions, the height of the free surface $y=\eta(x,t)$, the
turbulent energy density $k$, and its dissipation rate $\epsilon$.
The nonlinear model governing the evolution of the unknowns $\u$ and
$p$ is
\begin{eqnarray}
&&\F(p, \u)=\\ &&\left[ \begin{array}{c}
-u\pard{u}{x}-v\pard{u}{y}-\pard{p}{x}+g\sin\theta-\frac23\pard{k}{x}
+2\pard{~}{x}\left(\nu\pard{u}{x}\right)+\pard{~}{y}\left\{\nu\left(
\pard{u}{y}+\pard{v}{x}\right)\right\}\cr
-u\pard{v}{x}-v\pard{v}{y}-\pard{p}{y}-g\cos\theta-\frac23\pard{k}{y}
+2\pard{~}{y}\left(\nu\pard{v}{y}\right)+\pard{~}{x}\left\{\nu\left(
\pard{u}{y}+\pard{v}{x}\right)\right\}\cr
-u(x,\eta,t)\pard{\eta}{x}+v(x,\eta,t) \cr
-u\pard{k}{x}-v\pard{k}{y}+\left\{
\pard{~}{x}\left({\nu\over\sigma_k}\pard{k}{x}\right)+
\pard{~}{y}\left({\nu\over\sigma_k}\pard{k}{y}\right)\right\}+P_h
-\epsilon\cr -u\pard{\epsilon}{x}-v\pard{\epsilon}{y}+ \left\{
\pard{~}{x}\left({\nu\over\sigma_\epsilon}\pard{\epsilon}{x}\right)+
\pard{~}{y}\left({\nu\over\sigma_\epsilon}\pard{\epsilon}{y}\right)\right\}+
\frac{\Ceo}{T}P_h
-{\Cet}{\epsilon^2\over k} \end{array} \right].
\nonumber
\end{eqnarray} \esub
Here the eddy viscosity
 \begin{equation}
	 \nu=C_{\mu}{k^2\over \epsilon}
	 \label{nudef}
 \end{equation}
is a result of the turbulent mixing and varies in space and time. Later we
use the following approximate values of the constants
\begin{equation}\label{kevals}
   C_\mu=0.09\,,
   \quad\sigma_k=1\,,
   \quad\sigma_\epsilon=1.3\,,
   \quad{\Ceo}=1.44\,,
   \quad{\Cet}=1.92\,,
\end{equation}
in order to form definite models.
Also
\begin{displaymath}
	P_h=\nu\left[2\left(\pard{u}{x}\right)^2+2\left(\pard{v}{y}\right)^2
	+\left(\pard{u}{y}+\pard{v}{x}\right)^2\right]
\end{displaymath}
describes the generation of turbulence through instabilities associated
with mean-velocity gradients. $T$ is the time-scale of the turbulent eddies
and is frequently defined to be
\begin{displaymath}
	T=\max\left\{{k}/{\epsilon},
	C_T\sqrt{{\nu_m}/{\epsilon}} \right\}\,;
\end{displaymath}
the cut-off at viscous time-scales $\sqrt{{\nu_m}/{\epsilon}}$ is to
avoid a singularity in turbulent production at a wall, see
\cite[p470]{Durbin93b} for example.
However, here we eschew the incorporation of direct viscous effects
and so avoid this singularity by using $T=\avk/\ave$ as the typical
turbulent time-scale, where the overbars denote depth averages.
The downward slope of the bottom $\theta$ is assumed to be small and
to have negligible variation.

The boundary conditions on the bottom and the free surface are
important in the details of the construction of the low-dimensional model.
The following arguments lead to the given boundary conditions.
\begin{itemize}
\item The standard condition is that, in view of the extremely low density
of air, the pressure of the air on the fluid surface is effectively
constant which we take to be zero without loss of generality. Thus the
normal stress of the fluid across the free surface should vanish
\begin{equation}\label{BCp}
p+\frac{2}{3}k-\frac{2\nu}{1+\eta_x^2}\left[ \pard vy+\pard ux
-\eta_x\left( \pard vx+\pard uy\right)\right]=0\quad\mbox{on
$y=\eta$\,.}
\end{equation}
This is only approximately true---corrections should exist of the
order of $\overline{p'\eta'}$, and similarly for other equations
involving the free-surface.
However, the time-scale of gravity waves, $\sqrt{\ell/(2\pi g)}$,
associated with the turbulent length-scale, $\ell\propto
k^{3/2}/\epsilon$, should be typically much shorter than the turbulent
eddy turn-over time, $\ell/\sqrt{k}$ (true for the scaling introduced
in \S\S\ref{ss32}) and, as in \cite[\S\S2.3]{Hodges99}, we expect
there to be little interaction between the turbulent fluctuations and
the free surface dynamics.

\item In this work we assume that the horizontal extent of the flood
waves is small enough so that wind stress is negligible.
This is in contrast to large-scale geophysical simulations, such as
that by Arnold \& Noye \cite{AnNo82}, where the wind stress is very
important.
Thus the fluid surface is free of tangential stress:
\begin{equation}\label{FSBCu}
\left(1-\eta_x^2\right)\left(\pard uy+\pard vx\right)
+2\eta_x\left(\pard vy-\pard ux\right)=0\quad\mbox{on $y=\eta$\,.}
\end{equation}
A wind stress could be incorporated into the model by appropriately
replacing the zero right-hand side.
\item Symmetry conditions for $k$ and $\epsilon$ on the free surface
(see Arnold \& Noye~\cite{AnNo84} or Fredsoe \&
Deigaard~\cite[p117]{Fredsoe92} for examples) lead to
\begin{eqnarray}
\label{FSBCk}    \pard{k}{n}&=&0\quad\mbox{on $y=\eta$}\,, \\
\label{FSBCe} \mbox{and}\quad\pard{\epsilon}{n}&=&0\quad\mbox{on
$y=\eta$}\,.
\end{eqnarray}
These assert that the energy in turbulent eddies cannot be lost
or gained by transport through the free surface, and similarly for the
turbulent dissipation.
More sophisticated models of the free-surface effect upon turbulence
by Gibson \& Rodi \cite[p238]{Gibson89} use these zero net flux
boundary conditions.
Spilling breakers on the water surface could perhaps be modelled by a
turbulent production term on the right-hand side of these boundary
conditions.

\item At the ground, $y=0$, there must be no flow across the flat
bottom:
\begin{equation}\label{BoBCv}
v(x,0,t)=0\,.
\end{equation}

\item Other boundary conditions on the ground are more arguable
(compare our treatment with that of Arnold \& Noye
\cite{AnNo84,AnNo86}).
We are only interested in the flow outside of any molecular boundary
layer that may exist on the stream bed---we imagine that the structure
of any ordered viscous layer will be broken up by the roughness.
This is supported by recent experiments by Krogstad \& Antonia
\cite{Krogstad94} who show that roughness of a wall, even on a scale
1/100th the thickness of a turbulent boundary layer, tends to
\emph{reduce} the overall anisotropy of the turbulence.
A major limitation of the {\keps} model is the high level of
anisotropy near a wall, so such a reduction in anisotropy due to
roughness will favour the {\keps} model.
Note that we treat the ground as $y=0$ in the mathematical model even
though we imagine it to be rough.
In effect, ensemble means are also done over all realisations of a
``rough'' bed with mean slope $\theta$ and hence mean position $y=0$.

We suppose that the bottom inhibits the turbulence in its immediate
neighbourhood so that the turbulent energy falls to zero:
\begin{equation}
	k=0 \quad\mbox{on $y=0$}\,.
	\label{BoBCk0}
\end{equation}
In using the {\keps} model for near-wall turbulence,
Durbin~\cite{Durbin91,Durbin93} asserts that $\partial k/\partial y=0$
on the wall as well.
However, Figure~1 by Durbin~\cite{Durbin91} shows this latter
condition is only significant for the viscous boundary layer---a layer
we ignore due to the roughness of the ground.
Instead of this requirement, we place the fairly weak constraint on
the turbulent dissipation:
\begin{equation}
	\nu\frac{\partial\epsilon}{\partial y}\to 0 \quad\mbox{as $y\to 0$}\,.
	\label{BoBCe0}
\end{equation}
This is weak because $\nu\propto k^2\to 0$ as $y\to 0$. In essence, this
condition asserts that the bed does not directly act as a source or sink of
turbulent dissipation.

\item
Although $\nu\propto k^2\to 0$ as $y\to 0$, we suppose that $\nu$
approaches zero slowly enough so that turbulence is still an effective
mixing mechanism near the bed.
Thus, the ensemble mean horizontal velocity should also vanish on the
bed:
\begin{equation}
	u=0\quad\mbox{on $y=0$}\,.
	\label{BoBCu}
\end{equation}
These three boundary conditions on the bed are the same as those used
by ``low-Reynolds'' $k$-$\epsilon$ turbulent
models~\cite[p64]{Mohammadi94}.
The difference here is that we do not include the near wall dependence
upon local Reynolds numbers $R_t=k^2/\epsilon\nu_m$ and $R_y=\sqrt k
y/\nu_m$ because these involve molecular viscosity $\nu_m$.

The boundary conditions are different to those we used in an earlier
treatment of this problem~\cite{Mei94}.
The difference occurs because there we assumed that the stress
$\partial u/\partial y$ is small near the bottom and is more
appropriate to weakly turbulent flows.
Here we seek the dynamics of flows with a strong level of
turbulence leading to the boundary condition~(\ref{BoBCu}).
\end{itemize}

\section{Basis of the low-dimensional model}
\label{slowd}

In this paper we consider flows that vary ``slowly'' in the $x$ and
$t$ directions.
In this context, the meaning of ``slow'' is that the dynamics are
slower relative to the vertical mixing time induced by the turbulence.
In particular, the derivatives of the flow variables with respect to
$x$ and $t$ are small quantities that can be collected with the
``nonlinear'' part of the equations and treated as perturbations.
Hence we rewrite the equations as
\begin{eqnarray}\nonumber
\left[\matrix{0\cr \dot u\cr \dot v\cr \dot\eta\cr \dot k\cr
\dot\epsilon}\right] &=& \left[\matrix{0 &0 &\pard{~}{y} & 0 & 0 &0\cr
0 &\pard{~}{y}\left(\nu\pard{~}{y}\right) & 0 & 0 & 0& 0\cr
-\pard{~}{y} & 0&2\pard{~}{y}\left(\nu\pard{~}{y}\right) & 0
&-\frac23\pard{~}{y}& 0\cr 0 & 0&0&0&0&0\cr 0 &
0&0&0&{1\over\sigma_k}\pard{~}{y}\left(\nu\pard{~}{y}\right) &0\cr 0 &
0&0&0&0&
{1\over\sigma_\epsilon}\pard{~}{y}\left(\nu\pard{~}{y}\right)}\right]
\left[\matrix{p\cr u\cr v\cr \eta\cr k\cr \epsilon}\right] \\
\nonumber&& +\left[\begin{array}{c} \pard{u}{x} \\
-u\pard{u}{x}-v\pard{u}{y}-\pard{p}{x}+g\sin\theta-\frac23\pard{k}{x}
+2\pard{}{x}\left(\nu\pard{u}{x}\right)+\pard{}{y}
\left(\nu\pard{v}{x}\right)
\\
-u\pard{v}{x}-v\pard{v}{y}-g\cos\theta+\pard{}{x}\left[\nu\left(
\pard{u}{y}+\pard{v}{x}\right)\right]
\\
-u(x,\eta,t)\pard{\eta}{x}+v(x,\eta,t)
\\
-u\pard{k}{x}-v\pard{k}{y}+
\pard{}{x}\left({\nu\over\sigma_k}\pard{k}{x}\right)
+P_h-\lambda\epsilon
\\
-u\pard{\epsilon}{x}-v\pard{\epsilon}{y}+
\pard{}{x}\left({\nu\over\sigma_\epsilon}\pard{\epsilon}{x}\right)
+ \frac{\Ceo}{T}P_h
-{\Cet}\lambda{\epsilon^2\over k}
\end{array}\right]
\\  \label{lneqn}
&=& \cL(p, \u) +\cF(p,\u,\lambda)\,.
\end{eqnarray}
Treating the time and horizontal-space variations and the
``nonlinear'' terms as small, one sees that $\cL(p,\u)$ comprises the
leading term in the equation.

With a little adaptation, the operator $\cL$ has a critical space of
equilibrium points parametrised by the water depth $\eta$ and the
mean fields $\avu$, $\avk$ and $\ave$.
To ensure that the turbulent energy and turbulent dissipation are
critical modes of $\cL$ and thus retained in our model of turbulent
floods, we need $k$ and $\epsilon$ to be conserved to leading order.
The parameter $\lambda\in[0,1]$ is an artificial parameter which we
use to adjust the rate of decay of turbulent energy and its
dissipation; by making $\lambda$ small we initially neglect the
natural turbulent decay, but when $\lambda=1$ we recover the standard
$k$-$\epsilon$ model.
This is reasonable because the combined effect of $\dot k=-\epsilon$
and $\dot\epsilon=-\Cet \epsilon^2/k$ is a {\em slow} algebraic decay
that is appropriate for the long-term centre manifold dynamics.

\subsection{Vertical mixing}

The operator $\cL$ can be considered as determining the dominant
features of the evolution of the $k$-$\epsilon$ model~(\ref{EAKeFW}). It in
turn is primarily composed of the differential operator
\begin{displaymath}
     \pard{~}{y}\left(\nu\pard{~}{y}\right)\,,
\end{displaymath}
which represents vertical mixing by turbulent eddies.
By identifying this as the dominant term in the equations we are
physically supposing that turbulent mixing is stronger than the
other processes that redistribute momentum and turbulent energy.

The boundary conditions on the bottom~(\ref{BoBCk0})
and~(\ref{BoBCe0}) in conjunction with the $k$ and $\epsilon$
components of $\cL$ in~(\ref{lneqn}) admit homogeneous solutions $
k\propto \cbrt y$ and $\epsilon$ constant.
Such a cube-root profile in the vertical fits with arguments that the
turbulence should be weaker near the bottom due to its constraining
effects.
Given such a profile, the turbulent diffusivity $\nu\propto y^{2/3}$
and so the horizontal velocity component of $\cL$ also admits
homogeneous solutions $u\propto \cbrt y$.
Although our long-wave model will be expressed in terms of
depth-averaged quantities, we base the vertical structure that they
measure to be these cube-root profiles.

Traditionally, many theoretical approaches have assumed constant or
near constant vertical profiles (see \cite[\S\S4.3.1]{Fredsoe92} or
\cite[p670]{Prokopiou91b} for examples), as indeed we have also in an
earlier treatment of this problem \cite{Mei94}.
However, as seen in experiments the horizontal velocity profile is
typically curved (see \cite[Fig.~12]{Shiono91} or \cite{Bertschy83})
as is the turbulent energy (see \cite[Fig.~4.25]{Fredsoe92}).
A logarithmic profile is a well established approximation; here we
work with the cube-root profile as it is compatible with the \keps\
equations, is analytically tractable, and is a rough initial
approximation to the logarithmic profile.

These cube-root profiles result in downwards turbulent transport of
momentum and turbulent energy with constant flux and eventual removal
from the fluid at the bed.  In order to maintain flow at leading order
we modify the conservative free-surface boundary
conditions~(\ref{FSBCu}--\ref{FSBCk}) to supply the requisite flux at
leading order, and to remove the supply at higher order.  We
replace~(\ref{FSBCu}--\ref{FSBCk}) with the boundary conditions that
on $y=\eta$
\begin{eqnarray}
	(1-a\gamma)\left[
	\left(1-\eta_x^2\right)\left(\pard uy+\pard vx\right)
          +2\eta_x\left(\pard vy-\pard ux\right)
          \right] & =&\frac{1-\gamma}{3\eta}u
\,,
	\label{FSBCu1} \\
	(1-a\gamma)\left[
	\pard ky-\pard\eta x\pard k x
	\right] & =&\frac{1-\gamma}{3\eta}k
\,.
	\label{FSBCk1}
\end{eqnarray}
where the artificial parameter $\gamma$, as for $\lambda$ introduced
earlier, is small in the asymptotic scheme, but eventually will be set
to $1$ to recover the desired boundary
conditions~(\ref{FSBCu}--\ref{FSBCk}).
Such manipulation of the governing equations has worked well in
developing models of the laminar viscous flow of a thin fluid layer
\cite{Roberts94c,Roberts96b}.

The centre manifold analysis forms a power series in $\gamma$ which
needs to be summed for $\gamma=1$.
The parameter $a\neq 1$ is free for us to choose.
Initially we omitted $a$, equivalent to choosing $a=0$, but after
2~years of exploration we determined that an Euler transformation of
the series' in $\gamma$ greatly improves the convergence, e.g.~see
Hinch \cite[Chapt.~8]{Hinch91}.
Introducing $a\neq 0$ is equivalent to such an Euler transformation.
Another view of $a\neq 0$ is that of an over-relaxation parameter in
an iterative scheme.
In this problem it appears that $a\approx 1/2$ is a good compromise
between conflicting influences and is used henceforth.

From the special structure of $\cL$ with the modified boundary
conditions we deduce that there are four critical modes of interest in
the long-term dynamics.
These modes correspond to the leading-order conservation, and hence
long-life, of fluid mass, momentum, turbulent energy and turbulent
dissipation.
These modes span the space $\cM_0$
\begin{equation}
	(p,\u)=\left(g(\eta-y),\,U\cbrty,\,0,\,H,\,K\cbrty,\,E\right)\,,
    \label{Ecm0}
\end{equation}
where $U$, $H$, $K$ and $E$ are arbitrary functions of $x$ and $t$.
Note that the turbulent diffusivity, $\nu$, then also varies slowly in
$x$ and $t$; to leading order it is
\[
   \nu_0=C_\mu{16K^2\over 9E}\left(\frac{y}{\eta}\right)^{2/3}\,.
\]

As proven in Appendix~\ref{Scom}, the dominant operator $\cL$,
linearised about the space of equilibrium points, has eigenvalues which
are all negative (due to the decaying dynamics of turbulent
dispersion), except for the four zero eigenvalues corresponding to the
four conserved modes.
Since all other modes decay exponentially quickly, the long time
behaviour of the flow is determined by the functions $U(x,t)$,
$H(x,t)$, $K(x,t)$ and $E(x,t)$.
Respectively, these represent the vertically averaged horizontal
velocity, the surface elevation, the vertically averaged turbulent
energy, and the vertically averaged turbulent dissipation.
In essence, we construct a ``vertically averaged'' model, but in $u$
and $k$ there is structure in the vertical, roughly proportional to
$\cbrt y$, whose amplitude we measure by the vertical average.

\subsection{Approximating the centre manifold}
\label{ss32}

Centre manifold techniques systematically develop such a model in the
vertically averaged quantities.
Based on the relatively low-dimensional space of exponentially
attractive equilibria $\cM_0$, centre manifold theory \cite{Carr81}
suggests that the nonlinear terms ``bend'' $\cM_0$ to a nearby
manifold $\cM$ of slow evolution.
Further, $\cM$ will similarly attract exponentially quickly all
solutions in its vicinity; in standard formulations $\cM$
is called the centre manifold.
Once on $\cM$, solutions evolve slowly according to a low-dimensional
system of evolution equations---these evolution equations form the
simplified model of the original dynamics.
This general approach to forming low-dimensional models of dynamical
systems is reviewed by Roberts \cite{Roberts97a}.

In this problem, $\cM$ is parametrised by the four ``amplitudes,''
$U$, $H$, $K$ and $E$, which are functions of $x$ and evolve in time.
Due to the difficult nature of the nonlinear terms in the
$k$-$\epsilon$ model~(\ref{EAKeFW}) we have to be very careful about
these amplitudes and their derivatives.
We introduce two independent small parameters: $\delta$, as an
amplitude scale; and $\deltax$ to scale spatial derivatives.
Then we treat
\begin{equation}
u,k=\Ord{\delta^2}\,,\quad
\epsilon=\Ord{\delta^3}\,,\quad
\eta=h+\Ord{\delta^2}\,, \quad
\mbox{and}\quad
\pard{~}{x}=\Ord{\delta\deltax}\,.
\label{deltadefn}
\end{equation}
Note that in this scaling, the turbulent length-scale $\ell\propto
k^{3/2}/\epsilon=\Ord{1}$.
This ensures that the turbulent eddies modelled by $k$ and $\epsilon$
are not asymptotically larger than the water depth; large-scale
horizontal eddies are resolved by variations in the amplitudes of the
model.
Also from these scalings, the turbulent diffusivity
$\nu=\Ord{k^2/\epsilon}=\Ord{\delta}$ and thus the time-scale of
vertical mixing is $\Ord{1/\delta}$.
Consequently, we must consider horizontal scales larger than
$\Ord{1/\delta}$, which accounts for requiring the product
$\delta\deltax$ in the scaling of horizontal derivatives.
In standard applications of centre manifold theory, we are free to
scale the amplitudes in any reasonable fashion or indeed to treat the
amplitudes as independent; as discussed in \cite{Roberts88a} a change
in the scaling just reorders the appearance of the same set of terms
in the model.
In this application of centre manifold techniques, physical
considerations and the non-standard nonlinearities place the
constraint on the scaling that $\pard{~}{x}=\Ord{\delta\deltax}$ and
that $\theta=\Ord{\delta^3\deltax}$.
Nonetheless, we exploit usefully some of the capability of centre
manifold techniques to treat amplitudes as independent variables by
the above introduction of two independent small parameters, $\delta$
and $\deltax$.

Two other independent small parameters in this problem are: $\gamma$,
the artificial forcing at the free surface; and $\lambda$, an
artificial adjustment of turbulent interaction.
We treat all four of these parameters as independently small.

In terms of the field variables, we define the four amplitudes $U$,
$H$, $K$ and $E$ in terms of physical quantities such that
\bsub\label{ampdef}
\begin{eqnarray}
	\avu & = & \delta^2 U\,,
	\label{ampdefu} \\
	\eta & = & h+\delta^2 H\,,
	\label{ampdefh} \\
	\avk & = & \delta^2 K\,,
	\label{ampdefk} \\
	\ave & = & \delta^3 E\,,
	\label{ampdefe}
\end{eqnarray}
\esub
where the overbar denotes a vertical average over the whole fluid depth
at any $x$ and $t$, for example:
\begin{equation}\label{mean-u}
\avu=\frac1\eta\int_0^\eta u\,dy\,.
\end{equation}

Denoting the collective amplitudes by ${\bs(x,t)}=(U,H,K,E)$, pose the
low-dimensional assumption that the evolution of the physical
variables may be expressed in terms of the evolution of the four
amplitudes (effectively equivalent to the ``slaving'' principle of
synergetics \cite{Haken83}):
\begin{equation}
(p,\u)=\cV(y,\bs) \quad\mbox{such
that}\quad \pard{\bs}{t}=\cG(\bs)\,.
\label{CenSys}
\end{equation}
In general, we cannot find these functions $\cV$ and $\cG$ exactly
as this would be tantamount to solving exactly the original equations.
Instead we determine asymptotic approximations in the five small
parameters.

It would be decidedly awkward to explicitly write out an asymptotic
expansion in the five parameters.
But it is also inappropriate to link their relative magnitudes into
one parameter as we need to find relatively high-order in $\gamma$ but
not in the others.
Thus we apply an iterative algorithm in computer algebra to find the
centre manifold and the evolution thereon which is based directly upon
the Approximation Theorem~3 in \cite{Carr81,Roberts96a} and its
variants, as explained in detail by Roberts~\cite{Roberts96a}.
An outline of the procedure follows.

The aim is to find the functional $\cV$ and evolution $\cG$ such that
the pressure, velocity and turbulence fields described
by~(\ref{CenSys}) form actual solutions of the scaled turbulent
equations---this ensures fidelity between our model and the fluid
dynamics of the \keps~equations.
Suppose that at some stage in an iterative scheme we have some
approximation, $\tilde{\cV}$ and $\tilde{\cG}$.
We then seek a correction, $\cV'$ and $\cG'$, to obtain a more
accurate solution to the turbulence equations.
Substituting
\begin{displaymath}
(p,\u)=\tilde\cV+\cV'
\quad\mbox{such that}\quad
\pard{\bs}{t}=\tilde\cG+\cG'
\end{displaymath}
into the scaled turbulence equations then rearranging, dropping
products of corrections, and using a leading-order approximation
wherever factors multiply corrections (see \cite{Roberts96a} for
details), we obtain a system of equations for the corrections which is
of the form
\begin{equation}
        \cL\cV'=\tilde{\vec R}+\cE\cG'\,,
        \label{itform}
\end{equation}
where $\cE=\partial\cV/\partial\bs|_{\bs=\vec 0}$ is the basis for the
linear subspace $\cM_0$ in~(\ref{Ecm0}), and, most importantly,
$\tilde{\vec R}$ is the residual of the scaled \keps\ equations using
the current approximation $\tilde\cV$ and $\tilde\cG$.
This homological equation is solved by choosing corrections $\cG'$
to the evolution so that the right-hand side is in the range of $\cL$,
then the correction to the fields $\cV'$ is determined.
The solution is made unique by requiring that the amplitudes $\bs$
have some specific physical meaning, here the vertical averages of the
fields as in~(\ref{ampdef}).
Then the current approximation $\tilde{\cV}$ and $\tilde\cG$ is
updated.
The iteration is repeated until the residual of the governing
equations, $\tilde{\vec R}$, becomes zero to some order of error,
whence the centre manifold model will be accurate to the same order of
error (by the Approximation Theorem~3 of Carr~\cite{Carr81}).

A computer algebra program\footnote{The computer algebra package
{\reduce} was used because of its flexible ``operator'' facility.
At the time of writing, information about {\reduce} was available from
Anthony C.\ Hearn, RAND, Santa Monica, CA~90407-2138, USA.
\protect\url{mailto:reduce@rand.org}} was written to perform all the
necessary detailed algebra for this physical problem.
A listing is given in Appendix~\ref{Salg}.
The very important feature of this iteration scheme is that it is
performed until the residuals of the actual governing equations are
zero to some order of error.
Thus the correctness of the results that we present here is based only
upon the correct evaluation of the residuals and upon sufficient
iterations to drive these to zero.
The key to the correctness of the results produced by the computer
program is the proper coding of the \keps~turbulence equations.
These can be seen in the computed residuals within the iterative loop
of the program.

\section{Constructing the low-dimensional model}
\label{sss-homo}

As a first step in constructing a dynamical model we discard any
variation in $x$ and any influence of slope $\theta$.
Thus we first examine the dynamics of a uniform layer of turbulent
fluid, $k$ and $\epsilon\neq 0$, slowly decelerating, $u\neq 0$, due
to turbulent drag on the bed.

\subsection{The physical fields to low-order}

As discussed in \S\S\ref{ss32} on the vertical mixing operator, the
leading order approximation to the shape of the centre manifold is
just the solutions to $\cL\cV=0$.  We deduce
\begin{eqnarray*} && u\approx
\delta^2U(x,t)\cbrty\,,\quad v\approx 0\,,\quad k\approx
\delta^2K(x,t)\cbrty\,,\quad \\&& \epsilon\approx\delta^3E(x,t)\,,
\quad\mbox{and}\quad
\nu\approx\delta\Cm\frac{16K^2}{9E}\left(\frac{y}{\eta}\right)^{2/3}
\,.
\end{eqnarray*}
At higher orders in the small parameters $\delta$,
$\deltax$, $\lambda$ and $\gamma$ we construct more refined
descriptions of the fluid flow and its dynamics through the evolution
of the amplitudes.  However, we leave the influence of spatial
variations through non-zero $\deltax$ and $\theta$ until the next
section.

By iterations of the scheme outlined in the previous section we obtain
a basic description of the turbulence production and decay.
The nonlinear processes and boundary condition corrections modify the
cube-root profile and simultaneously determine the slow evolution of
the amplitudes.

It is useful to record the asymptotic expansions directly in terms of
physical quantities $\eta$, $\avu$, $\avk$ and $\ave$ rather than the
corresponding artificially scaled quantities $H$, $U$, $K$ and $E$.
We find the following expressions for the first significant
modifications to the fields within the fluid, written in terms of a
scaled vertical coordinate $\zeta=y/\eta$ which ranges from $\zeta=0$
at the bed to $\zeta=1$ at the fluid surface:
\bsub\label{mdl0v}
\begin{eqnarray}
	v & \approx & \label{mdl0vv}
	      0
	      \,,  \\
	u & \approx & \label{mdl0vu}
	       \vz_0(\zeta)\avu
+\left[\Ceo\se\vz_1(\zeta)-\sk\vz_2(\zeta)\right]\frac{\avu^3}{\avk}
\nonumber\\&&\quad{}
	       +\left(\Cet\se-\sk\right)\vz_3(\zeta)
	            \frac{\lambda\eta^2\avu\ave}{\nut\avk}
	       \,,\\
	k & \approx & \label{mdl0vk}
	       \vz_0(\zeta)\avk
	       +\left[\Ceo\se\vz_1(\zeta)+\sk\vz_2(\zeta)\right]\avu^2
	       +\left(\Cet\se+\sk\right)\vz_3(\zeta)
	            \frac{\lambda\eta^2\ave}{\nut}
	       \,,  \\
	\epsilon & \approx &   \label{mdl0ve}
	       \ave
	       +\Ceo\se\epsilon_{p}(\zeta)\frac{\avu^2\ave}{\avk}
	       +\Cet\se\epsilon_{d}(\zeta)\frac{\lambda\eta^2\ave^2}{\nut\avk}
	       \,, \\
\nu&\approx&\nu_0(\zeta)\frac{{\bar{k}}^2}{\bar{\epsilon}}
+\left[-\Ceo\se\nu_1(\zeta)+\sk\nu_2(\zeta)\right]\frac{\ave\eta^2}{\avk}
\nonumber\\&&\quad{}
+\left[\Cet\se\nu_3(\zeta)
-\sk\nu_4(\zeta)\right]\frac{\ave^3\eta^4}{\avk^4}\,.
\label{mdl0vn}
\end{eqnarray}
\esub
These expressions are correct to errors
$\Ord{\delta^6+\lambda^3+\gamma^3,\deltax}$ where, for example, a
multinomial term
\begin{displaymath}
	\delta^a\lambda^b\gamma^c\deltax^d =
	\Ord{\delta^A+\lambda^B+\gamma^C,\deltax^D}
	\quad\mbox{if}\quad
	\frac{a}{A}+\frac{b}{B}+\frac{c}{C}\geq 1
	\,,\ \mbox{and}\
	d\geq D\,.
\end{displaymath}
The vertical structure functions occurring in the expressions on the
right-hand side of~(\ref{mdl0v}) are as follows.
\begin{itemize}
\item  For the turbulent dissipation:
\begin{eqnarray*}
	\epsilon_{p}(\zeta) & = &
 \frac{4}{9}\zeta ^{4/3}-\frac{8}{9}\zeta^{2/3}
 +\frac{12}{35}
     \,,  \\
	\epsilon_{d}(\zeta)&=& -\frac{243}{512}\zeta ^{4/3}
	+\frac{81}{128}\zeta-\frac{405}{3584}
	\,,
\end{eqnarray*}
as shown in Figure~\ref{figve}.
\begin{figure}
	\centering
	\includegraphics{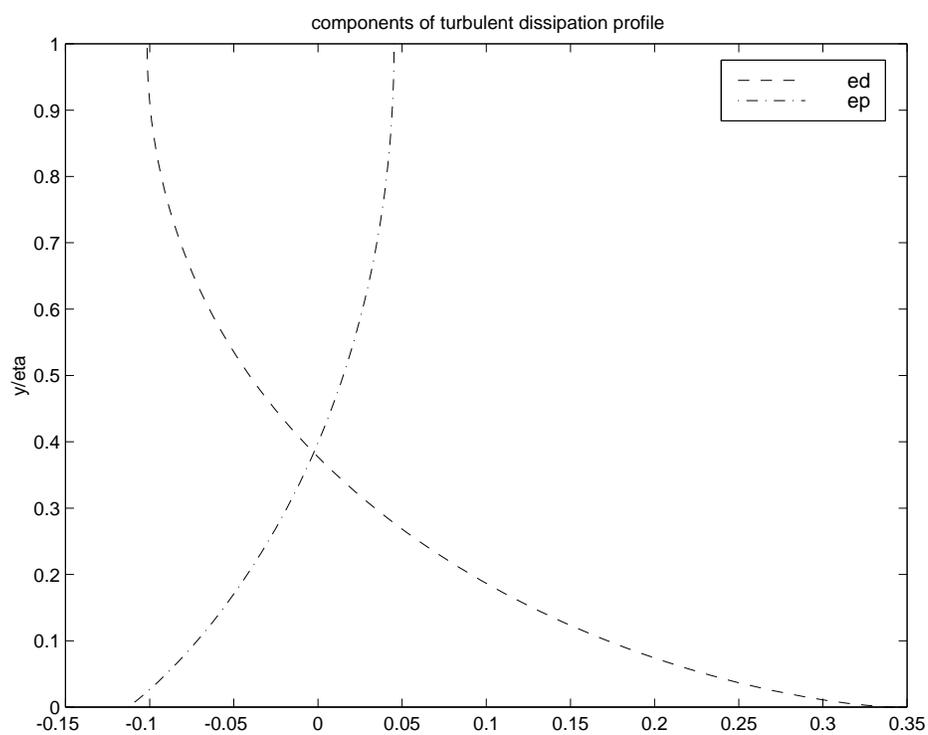}
	\caption{the vertical structure of the turbulent dissipation field
within the fluid as a function of the scaled vertical coordinate
$\zeta=y/\eta$: --$\cdot$--$\cdot$--$\cdot$--, $\epsilon_{p}(\zeta)$;
--~--~--~--, $\epsilon_{d}(\zeta)$.}
	\label{figve}
\end{figure}
See the effect of turbulent dissipation production, at a rate
proportional to $\avu^2$, through the velocity shear.
Since velocity shear is largest near the bed, as seen in the shape of
$\epsilon_{p}(\zeta)$ this enhances turbulent dissipation $\epsilon$
near the bed.

However, the natural turbulent dissipation within the fluid causes a
greater decay of turbulent dissipation near the bed, due to the
smaller turbulent energy there, and so counters this enhancement.
Being proportional to $1/\nut$, this effect on the $\epsilon$-profile
is greatest in weakly turbulent flows.

\item  For the turbulent energy density:
\begin{eqnarray*}
	\vz_0(\zeta) & = & \frac{4}{3}\zeta^{1/3}
	           +\gamma\left(\frac{1}{12}\zeta^{1/3}
	           -\frac{1}{6}\zeta^{5/3}\right)
	\,,  \\
	\vz_1(\zeta) & = &
 \frac{16}{135}\zeta^{5/3}-\frac{32}{135}\zeta +\frac{8}{81}\zeta^{1/3}
    \,,  \\
	\vz_2(\zeta) & = &
\frac{1}{9} \zeta ^{5/3}-\frac{4}{9} \zeta ^{2/3}
+\frac{3}{10} \zeta^{1/3}
\,,  \\
	\vz_3(\zeta)&=& -\frac{27}{256} \zeta ^{5/3}
	+\frac{9}{64} \zeta ^{4/3}
	-\frac{99}{3584} \zeta^{1/3}
	\,,
\end{eqnarray*}
as shown in Figure~\ref{figvk}.
\begin{figure}
	\centering
\includegraphics{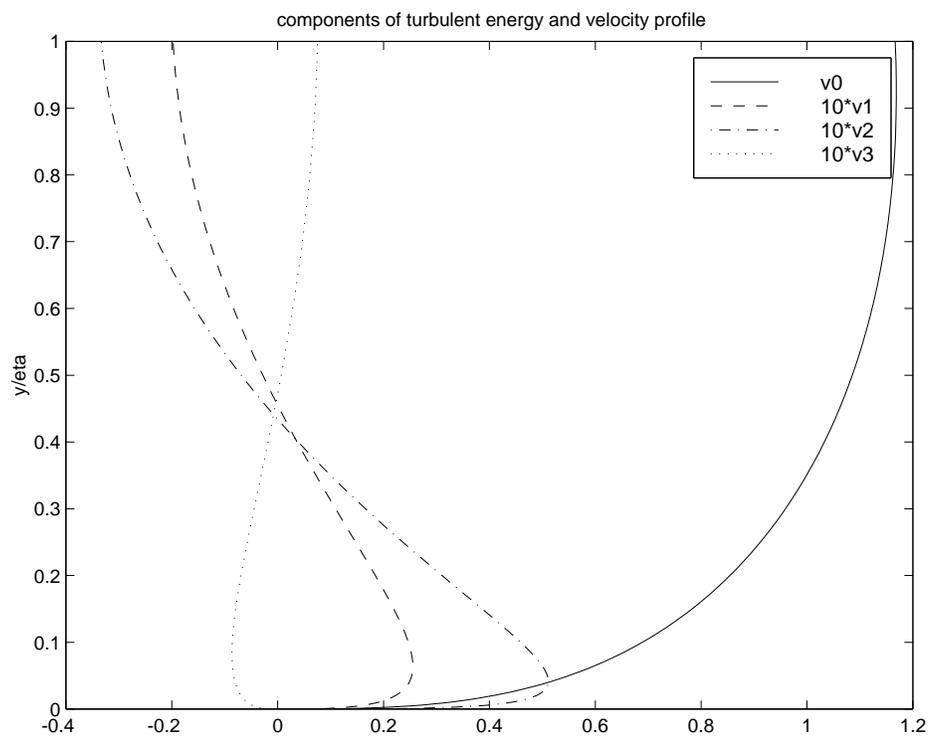} \caption{the vertical structure
functions for the turbulent energy density and horizontal velocity
fields within the fluid as a function of the scaled vertical
coordinate $\zeta=y/\eta$: --------, $\vz_0(\zeta)$ (with
$\gamma=1$); --~--~--, $10\times \vz_1(\zeta)$;
--$\cdot$--$\cdot$--$\cdot$--, $10\times \vz_2(\zeta)$; $\cdot
\cdot \cdot \cdot \cdot \cdot$, $10\times \vz_3(\zeta)$.  }
	\label{figvk}
\end{figure}
Observe the cube-root structure in the vertical is modified to
$\vz_0(\zeta)$.
As shown in Figure~\ref{figvk}, when $\gamma$ is set to $1$ to recover
the original boundary conditions from~(\ref{FSBCk1}), the cube-root
dependence is maintained near the bed, but is effectively flattened
near the fluid surface to closely approximate the absence of turbulent
energy flux through the free surface.
This correction is simultaneously determined with a corresponding
decay term in the evolution equations, as seen below, due to the
removal of the sustaining flux.
We look even closer at the effects of modifying the free-surface
boundary conditions in \S\S\ref{sss-conv}.

Also the effect of turbulent production, proportional to $\avu^2$,
through the velocity shear is largest near the bed; as seen in the
shape of $\vz_1(\zeta)$ and $\vz_2(\zeta)$ this enhances turbulent
energy $k$ near the bed.

However, the natural turbulent dissipation within the fluid causes a
relatively greater decay of turbulent energy near the bed as compared
with the body of the fluid, as seen in $\vz_3(\zeta)$, and so counters
this enhancement.  Being proportional to $1/\nut$, this effect on the
$k$-profile is greatest in weakly turbulent flows.

\item The basic cube-root structure of the horizontal velocity is
modified in exactly the same way and for the same reasons as for the
basic turbulent energy $k$-profile.

Modifications of the velocity profile due to the turbulent production
and dissipation occur, but they occur primarily through the indirect
effects of modifications to the turbulent diffusivity profile
$\nu(\zeta)$.
These are weak due to the subtractions in~(\ref{mdl0vu}).

\item The corresponding vertical structure of the turbulent mixing
coefficient $\nu$ is shown in Figure~\ref{fignuz} where the five
components are
\begin{eqnarray*}
\nu_0(\zeta)&=&\frac{20}{9}\zeta^{2/3}-\frac{8}{9}\zeta^2\,,
\\
\nu_1(\zeta)&=&\frac{64}{135}\zeta^2-\frac{128}{135}\zeta^{4/3}
+\frac{2944}{8505}\zeta^{2/3}\,,
\\
\nu_2(\zeta)&=&-\frac{32}{27}\zeta+\frac{4}{5}\zeta^{2/3}
     +\frac{8}{27}\zeta^2\,,
\\
\nu_3(\zeta)&=&\frac{57}{448}\zeta^{2/3}-\frac{3}{4}\zeta^{5/3}
+\frac{9}{16}\zeta^2\,,
\\
\nu_4(\zeta)&=&\frac{33}{224}\zeta^{2/3}+\frac{9}{16}\zeta^2
-\frac{3}{4}\zeta^{5/3}\,.
\end{eqnarray*}
\begin{figure}
	\centering
\includegraphics{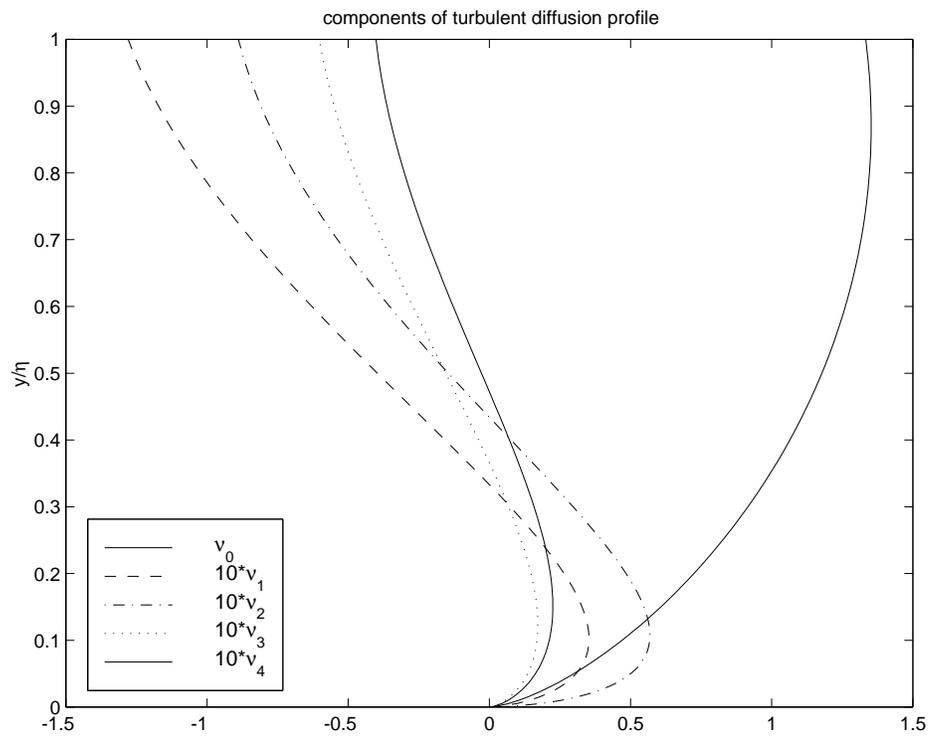}
\caption{the vertical structure functions for the turbulent mixing
coefficient within the fluid as a function of the scaled vertical
coordinate $\zeta=y/\eta$: -------- (right), $\nu_0(\zeta)$ (with
$\gamma=1$); --~--~--, $10\times \nu_1(\zeta)$;
--$\cdot$--$\cdot$--$\cdot$--, $10\times \nu_2(\zeta)$; $\cdot \cdot
\cdot \cdot \cdot \cdot$, $10\times \nu_3(\zeta)$; -------- (left),
$\nu_4(\zeta)$.  }
	\label{fignuz}
\end{figure}
\end{itemize}

Simultaneously with the determination of the above fields, the
solvability condition for the linear equations of the
form~(\ref{itform}) supplies terms in the asymptotic, low-dimensional
evolution equations for the amplitudes of the four critical modes.
Writing these in terms of physical variables we find, with errors
$\Ord{\delta^6+\lambda^3+\gamma^3,\deltax}$: \bsub\label{mdl0}
\begin{eqnarray}
	\frac{\partial \eta}{\partial t}&\sim&
	0\,,
	\label{mdl0h} \\
	\frac{\partial\avu}{\partial t}
	&\sim&
	-\frac{56\gamma\nut}{81\eta^2}\avu
	-\frac{\lambda}{16}\left[\Cet\se-\sigma_k\right]
	              \frac{\ave\avu}{\avk}
	+\frac{16}{81}\left[\frac{32}{45}\Ceo\se-\sigma_k\right]
	              \frac{\nut\avu^3}{\eta^2\avk}
	\,,
	\label{mdl0u} \\
	\frac{\partial\avk}{\partial t}&\sim&
	-\lambda\left[\frac{7}{8}+\frac{ \Cet \se}{16\sk}\right]\ave
	-\frac{56\gamma\nut}{81\sigma_k\eta^2}\avk
	+\frac{16}{243}\left[7+\frac{32\Ceo\se}{15\sigma_k}\right]
	   \frac{\nut}{\eta^2}\avu^2
	\,,
	\label{mdl0k} \\
	\frac{\partial\ave}{\partial t}&\sim&
	-\lambda \frac{9}{8}\Cet\frac{\ave^2}{\avk}
	+\frac{256}{243}\Ceo \frac{\nut\ave}{\eta^2\avk}\avu^2
	\,,
	\label{mdl0e}
\end{eqnarray}
\esub
where
\begin{equation}
	\nut(x,t)=\Cm\frac{\avk^2}{\ave}
	\label{mdl3nu}
\end{equation}
is a measure of the local turbulent diffusivity.  These form a crude
approximation to the evolution equations for the four amplitudes of
the model when there are no horizontal variations.

The two fifth-order evolution equations~(\ref{mdl0k}--\ref{mdl0e})
summarise the turbulent production and decay processes.  Setting the
artificial parameters $\lambda=\gamma=1$ to approximate the dynamics
of the original problem, observe: first, the natural decay of
turbulent energy and dissipation in the body of the fluid; second,
decay of turbulent energy with coefficient proportional to
$\nut/\eta^2$ via turbulent mixing transporting energy to the bed; and
third, the generation of turbulence energy and its dissipation through
the shear in the vertical, proportional to $\nut\avu^2/\eta^2$.

The horizontal velocity evolution~(\ref{mdl0u}) similarly includes
terms: $\nut\avu/\eta^2$ which represents the effective drag of the
bottom via turbulent mixing to the bed; and weak cubic, proportional
to $\nut\avu^3/(\eta^2\avk)$, and linear, proportional to
${\ave\avu}/{\avk}$, modification of this drag through changes in the
stress tensor in the momentum equations (note that the coefficients
are the difference of two terms, and that with the usual values for
the parameters~(\ref{kevals}) there is significant cancellation).

The free-surface stays horizontal, equation~(\ref{mdl0h}), because
there are no horizontal gradients until we look at order $\deltax$
effects in a later subsection.

\subsection{Convergence in the artificial parameters}
\label{sss-conv}

One limitation on the accuracy of the above model is that even within
the \keps~model of turbulence the coefficients are only approximate.
This is due to both the modification of the free surface boundary
conditions on $u$ and $k$ to~(\ref{FSBCu1}--\ref{FSBCk1}); and the
introduction of $\lambda$ in~(\ref{lneqn}).
Although setting $\gamma=1$ recovers the original boundary
conditions~(\ref{FSBCu}--\ref{FSBCk}) and $\lambda=1$ recovers the
original {\keps} model, there is no certainty that this will give a
model which is a good approximation to the ``true'' system.
In essence the coefficients in the model are multi-variable Taylor
series in $\gamma$ and in $\lambda$.
In this subsection we present evidence that these series converge
for $\gamma=\lambda=1$ and so we can form a reasonable model.

Arbitrarily high order terms in the centre manifold expansion may be
computed in principle.
Our computer algebra program currently is limited by memory and time
constraints to about 8th order in $\gamma$ and lower orders in other
parameters.\footnote{In some applications
\cite{Mercer90,Roberts93,Watt94b} such routine computations can be
performed to 30th order and are used to show convincingly the
convergence or otherwise of the series expansions.} This is only
attainable by the simplification of setting the \keps~parameters to
the conventional numerical values in~(\ref{kevals}).
By executing the \reduce{} program and discarding terms
$\Ord{\delta^6, \gamma^8, \lambda^2, \deltax}$, we discover more terms
in the series in $\gamma$.
Listed in Table~\ref{tabg} are the expansions of some of the
coefficients appearing later in the models.
\begin{table}
\caption{terms in the series expansions in $\gamma$ of selected
coefficients in the model for homogeneous turbulent decay.  The last
row is the sum of known terms at $\gamma=1$.}
	\protect\label{tabg}
	\begin{center}\begin{tabular}{|c||c|c|c||c|c|}
		\hline
		&\multicolumn{3}{c||}{in $\partial\avu/\partial t$}
		&\multicolumn{2}{c|}{in $\partial\avk/\partial t$}\\
		\hline
		  & $-\nut\avu/\eta^2$ & $-\nut\avu\lamb/\eta^2$
		  & $\nut\avu^3/\eta^2\avk$
		  &$\nut\avu^2/\eta^2$ & $-\ave$ \\
		\hline\hline
		$1$       &     0      & $+1.03889$ & $+0.06542$
		& $+0.72385$ & $+1.03100$ \\
		$\gamma  $& $+0.69136$ & $-0.86096$ & $+0.01388$
		& $-0.13682$ & $-0.05665$ \\
		$\gamma^2$& $+0.22783$ & $+0.14923$ & $-0.01564$
		& $-0.02193$ & $+0.01602$ \\
		$\gamma^3$& $+0.07479$ & $+0.02700$ & $-0.00921$
		& $+0.00569$ & $+0.00250$ \\
		$\gamma^4$& $+0.02448$ & $+0.00462$ & $-0.00333$
		& $+0.00814$ & $+0.00027$ \\
		$\gamma^5$& $+0.00801$ & $+0.00069$ & $-0.00080$
		& $+0.00549$ & $-0.00002$ \\
		$\gamma^6$& $+0.00262$ & $+0.00006$ & $-0.00003$
		& $+0.00300$ & $-0.00003$ \\
		$\gamma^7$& $-0.00086$ & $-0.00002$ & $+0.00017$
		& $+0.00147$ & $-0.00001$ \\
		\hline\hline
		$\sum$    & $\phantom{+}1.02995$   & $\phantom{+}0.35950$
		& $\phantom{+}0.05040$
		& $\phantom{+}0.58892$    & $\phantom{+}0.99307$  \\
		\hline
	\end{tabular}
	\end{center}
\end{table}

Look down the columns in the table and see that the coefficients in
each sequence generally decrease by at least a factor of two.
This suggests that the radii of convergence of the series in $\gamma$
are roughly two or more.
Thus simply evaluating the series' at $\gamma=1$ is reasonably
good---some are shown in the bottom line of the
table.\footnote{Actually, the introduction of the parameter $a$, and
the selection of $a=1/2$, was motivated by our original series in
$\gamma$ exhibiting singularities for $\gamma\approx -1$, as indicated
by Domb-Sykes plots \cite{Hinch91}.
These singularities ruined the convergence at $\gamma=1$.
However, an Euler transform of the series to accelerate convergence is
precisely equivalent to choosing non-zero $a$ and a few numerical
experiments lead to $a\approx1/2$ causing good convergence.}

The convergence in the parameter $\lambda$ is problematical because it
seems to appear always in the combination $\lambda\eta^2\ave
/(\nut\avk)$, that is $\lambda\eta^2\ave^2/\avk^3$.
Thus convergence depends upon the properties of the solution which is
generally unknown beforehand.
We suggest that truncating to linear terms in $\lambda$ forms an
adequate approximation.
It seems at least self-consistent to do this as later homogeneous
solutions, namely~(\ref{Edecay}--\ref{Evenant}), show a balance for the
relatively small value $\eta^2\ave^2/\avk^3\approx 0.2$.
Thus the nonlinear terms in $\lambda$ are generally expected to have
a negligible influence in most flows of interest.

First, the series' are summed for $\gamma=1$ as discussed in the
previous subsection.
Then introducing
\begin{equation}
	\lamb=\frac{\lambda\eta^2\ave^2}{\avk^3}
	=\Cm\lambda\frac{\eta^2/\nut}{\avk/\ave}
	\propto\lambda\frac{\mbox{vertical
	mixing time}}{\mbox{turbulent eddy time}}\,,
	\label{Elamb}
\end{equation}
for brevity we write the model of decay of homogeneous turbulent flow
as follows, with errors $\Ord{\delta^6, \lambda^2, \deltax}$,
\bsub\label{mdld}
\begin{eqnarray}
    \Dt\eta & = & 0\,, \\
	\Dt\avu & = &
	-(1.030+0.359\,\lamb)\frac{\nut\avu}{\eta^2}
	+(0.0504-0.243\,\lamb)\frac{\nut\avu^3}{\eta^2\avk}\,,
	\label{mdldu}\\
	\Dt\avk & = &
	-(0.0927+0.993\,\lamb)\frac{\avk^3}{\eta^2\ave}
	+(0.589+0.516\,\lamb)\frac{\nut\avu^2}{\eta^2}\,,
	\label{mdldk}\\
	\Dt\ave & = &
	-2.101\,\lamb\frac{\avk^2}{\eta^2}
	+(1.552-3.215\,\lamb)\frac{\nut\ave\avu^2}{\avk\eta^2}\,.
	\label{mdlde}
\end{eqnarray}
\esub Note that setting $\lambda=1$ to recover the original problem is
just equivalent to using $\lamb={\eta^2\ave^2}/{\avk^3}$.
Observe that the first terms on the right-hand sides of the above
represent decay terms through, for example in the $\avu$ and $\avk$
equations, turbulent transport to the stream bed.
The last terms in the $\avk$ and $\ave$ equations represent the
production of turbulence through the velocity shear.

One feature of the model derived here is that it has no adjustable
coefficients.
All constants are derived from well known physical parameters and accepted
constants of the $k$-$\epsilon$ equations.
Despite its relative complexity, the model has been systematically
derived and the constants which appear are well-defined.
However, there are adjustable parameters, namely the order of
truncation of the series expansions.
The model~(\ref{mdld}), for example, contains just the low-order terms
in expansions in $\delta$ and $\lambda$.

\subsection{Dynamics of spatial structure}
\label{sss-wavad}

The leading order effect of horizontal gradients, such as due to a
sloping free surface, is found by computing terms of order $\deltax$
in the asymptotic expansions.
We describe these in this subsection.

Dominantly, horizontal gradients affect the velocity and pressure
fields.
By computing terms to order $\delta^3\deltax$ we find that the
velocity fields~(\ref{mdl0vv}--\ref{mdl0vu}) are modified to \bsub
\begin{eqnarray}
	v  &=& \label{mdl1vv}
	      -\zeta^{4/3}\eta \pard{\avu}x
	      + \Ord{\delta^6+\lambda^3+\gamma^3+\deltax^3}\,,\\
	u &=& \label{mdl1vu}
	      \cdots
	      +3\vz_3(\zeta)\frac{g\eta^2}{\nut}\pard\eta x
	      +\Ord{\delta^6+\lambda^3+\gamma^3+\deltax^3}\,,
\end{eqnarray}
\esub where the $\cdots$ indicate the terms on the right-hand side
of~(\ref{mdl0vu}), and where $\vz_3(\zeta)$ is drawn in
Figure~\ref{figvk}.
The shape of $v$ is required by the continuity equation.
The modification to $u$ asserts reasonably that at low levels of
turbulence, large $1/\nut$, horizontal accelerations through
decreasing depth, $\eta_x<0$, cause the fluid to respond with a
flatter profile through a subtraction of $\vz_3(\zeta)$ from
$\vz_0(\zeta)$ as seen in Figure~\ref{figvk}.

The structure of the fields within the fluid rapidly become more
complicated at higher order.  We do not detail the fields any more.

By executing the computer algebra program and discarding generated
terms $\Ord{\delta^6, \gamma^6, \lambda^2, \deltax^2}$, we discover
first order effects of horizontal variations with sufficient terms in
the series in $\gamma$ to sum them reliably for $\gamma=1$.
We find the same production and decay terms identified in~(\ref{mdld})
and, in addition, extra terms in the horizontal gradients.
Using the accepted values~(\ref{kevals}) for the constants of the
{\keps} equations, we obtain the following model with our best
estimates of its coefficients: \bsub\label{mdl3}
\begin{eqnarray}
\label{mdl3h}
	\frac{\partial \eta}{\partial t}&\sim&
	- \frac{\partial(\eta\avu)}{\partial x}\,,
\\
\frac{\partial\avu}{\partial t}&\sim&
-(1.030+0.359\,\lamb)\frac{\nut\avu}{\eta^2}
+(0.0504-0.243\,\lamb)\frac{\nut\avu^3}{\eta^2\avk}
\nonumber \\ \nonumber &&{}
+\left[
0.961-0.019\,\lamb-(0.019-0.087\,\lamb) \frac{\avu^{2} }{\avk}
\right]
g\left(\theta-\frac{\partial\eta}{\partial x}\right)
\\ \nonumber &&{}
-(1.105+0.104\,\lamb)\avu \frac{\partial \avu}{\partial x}
-(0.032-0.056\,\lamb)\frac{\avu^{2}}{ \avk}\frac{\partial \avk}{\partial x}
\\&&\quad{}
+(0.025-0.041\,\lamb) \frac{\avu^{2}}{ \ave}\frac{\partial \ave}{\partial x}
\,,\label{mdl3u}\\
\frac{\partial\avk}{\partial t}&\sim&
-0.0927\frac{\avk^3}{\eta^2\ave}-0.993\,\ave
+(0.589+0.516\,\lamb)\frac{\nut\avu^2}{\eta^2}
\nonumber\\ \nonumber &&{}
-(0.025+0.011\,\lamb) g \avu\left(\theta
-\frac{\partial \eta}{\partial x}\right)
\\ \nonumber &&{}
-(1.106-0.065\,\lamb) \avu \frac{\partial \avk}{\partial x}
-(0.030+0.056\,\lamb)\avk\frac{\partial \avu}{\partial x}
\\&&\quad{}
+(0.025-0.060\,\lamb)\frac{\avu\avk}{\ave}\frac{\partial\ave}{\partial x}
\,,\label{mdl3k}\\
\frac{\partial\ave}{\partial t}&\sim&
-2.101\frac{\ave^2}{\avk}
+(1.552-3.215\,\lamb)\frac{\nut\ave\avu^2}{\avk\eta^2}
\nonumber\\ \nonumber &&{}
+\left(-0.006+0.562\,\lamb\right) g \frac{ \avu \ave}{\avk}\left(\theta
-\Dx\eta\right)
\\  &&{}
-0.173\,\lamb\ave\Dx\avu
+0.533\,\lamb\frac{\ave\avu}{\avk}\Dx\avk
-(1+0.735\,\lamb) \avu \frac{\partial \ave}{\partial x}
\,.\label{mdl3e}
\end{eqnarray}
\esub We expect that the coefficients in the above equations, when
considered as a model of the {\keps} equations given in
Section~\ref{skemod}, are accurate as shown.
Except for the surface equation~(\ref{mdl3h}), the first line in each
equation is the same as in the horizontally homogeneous
model~(\ref{mdl0}); subsequent lines detail the additional terms
needed to begin modelling long waves.  A simpler version of the above
model, obtained by omitting terms with small coefficients, is
recorded in the Introduction as the model~(\ref{mdli}).

Equation~(\ref{mdl3h}) is an exact statement of the conservation of water
and is not modified by any higher-order effects. To $\delta^3\deltax$,
it may be written
\bsub\label{mdl2}
\begin{equation}
	\frac{\partial\eta}{\partial t}+h\frac{\partial \avu}{\partial x}=0\,,
	\label{mdl2h}
\end{equation}
which is a linear description of the conservation of water.
Similarly, with $\theta=0$ and in very low levels of turbulence
($\nut\approx 0$) the horizontal momentum
equation~(\ref{mdl3u}) may be written to $\delta^3\deltax$ as
\begin{equation}
	\frac{\partial\avu}{\partial t}
	=-0.961\,g\frac{\partial \eta}{\partial x}\,.
	\label{mdl2u}
\end{equation}
\esub This describes the horizontal acceleration due to slope of the
fluid surface.
These last two coupled equations form a standard description of linear
wave dynamics except for one remarkable feature: the effect of gravity
is reduced by the factor $0.961$.
For example, this would predict that even low levels of
turbulence reduces the phase speed of waves by about two percent.
As in thin films of viscous fluid~\cite{Roberts94c}, the phenomenon is
due to the response of the fluid, approximately $v_0(\zeta)$ shown in
Figure~\ref{figvk}, being at an angle to the forcing $1$ (either due
to gravity or horizontal pressure gradients) when considered in the
space of functions on $[0,\eta]$.
Consequently, the forcing is less effective.
Such a depression in phase speed may be observable in the propagation
of long-waves on turbulent flow.\footnote{This modelling approach
shows that where there is vertical or cross-sectional structure,
depth-averaging or cross-sectional integration is generally unsound as
a modelling tool.
The reason is that it is the size and structure of the dynamical modes
which determine the evolution (here approximately cube-root), and not
the particular amplitudes used to measure the motion (here depth
averages).
}

Returning to the order $\delta^5$ momentum equation~(\ref{mdl3u}) we
note several interesting effects.
\begin{itemize}
	\item The first line contains the turbulent drag terms identified
	in the previous subsection.

 \item The second lines describe the effects of surface and bed slope.
 Within the square brackets:
	\begin{itemize}
		\item the first term gives the depression of wave speed
		discussed above;

 \item the second term very weakly enhances the phase speed correction
 in turbulent flow;

 \item whereas it is difficult to ascribe one definite cause to the last
term, coefficient modifications of the form $\avu^2/\avk$ are common in
this model and reflect the relative importance of the turbulence on the
mean flow.
\end{itemize}

\item The third and fourth lines are dominated by the nonlinear
advection term $\avu\avu_x$, with coefficient approximately $1.1$.
This coefficient is larger than 1 because of the shear: the maximum
$u(y)>\avu$ advects itself faster than $\avu$.
This third line also shows small ``cross-talk'' effects in the
advection through the $\avu^2\left(\log\avk\right)_x$ and
$\avu^2\left(\log\ave\right)_x$ terms.
\end{itemize}

The dynamics of $k$ and $\epsilon$ averages are given by
equations~(\ref{mdl3k}--\ref{mdl3e}).
\begin{itemize}
	\item  The first lines of each equation are the same turbulent
production and decay terms identified in the previous subsection.

 \item The next line in each equation may arise from the modification
of the turbulent production through the change of the velocity profile,
seen in~(\ref{mdl1vu}), due to horizontal acceleration.

 \item The remaining terms simply represent horizontal advection by the
fluid velocity. Note that different properties are advected at different
effective speeds\footnote{Though due to the nonlinear interaction terms we
should really report on the speeds associated with the characteristics of
the equations.} as indicated by the different coefficients of the
$\avu\partial/\partial x$ terms: $1.1$ for $\avu$ and $\avk$,
and $1$ for $\ave$.

\end{itemize}
The model~(\ref{mdli}) reported in the introduction is a simplified
version of~(\ref{mdl3}).
The solutions described in the next section show that the terms
neglected from~(\ref{mdl3}) in writing~(\ref{mdli}) are relatively
small, contributing at most a few percent in the numerical balance of
the terms, and so may be neglected at least for initial exploration.

\section{Predictions of the new model}
\label{seg}

In this section we investigate some of the predictions the newly
derived model~(\ref{mdl3}) might make.
We look at decaying turbulence, uniform flow on slopes, approximation
to the St~Venant equations, and a dam break simulation.

\subsection{Decaying turbulence}

Homogeneous turbulence decays algebraically.
If there is no slope ($\theta=0$), no variations in $x$, no mean flow
($\avu=0$) and no surface waves ($\eta=\mbox{const}$), then it is
consistent to seek solutions of the model~(\ref{mdl3}) in the form
$\avk\propto t^{-2}$ and $\ave\propto t^{-3}$.
Substituting and solving for the constants of proportionality the
model~(\ref{mdl3}) predicts the turbulence decays according to
\begin{equation}
	\avk \sim 8.97\,\eta^2t^{-2}\,,\quad
	\ave \sim 12.8\,\eta^2t^{-3}\,,\quad
	\nut \sim 0.565\,\eta^2t^{-1}\,,\quad
	\lamb\sim 0.227\,.
	\label{Edecay}
\end{equation}
for large time $t$.  The turbulence ultimately decays with the
balance $\ave\sim 0.48\,\avk^{3/2}/\eta$.

However, the transients towards this large time behaviour may be long.
There are two regimes of interest characterised by large and small
$\lamb$ compared to $0.227$ (recall from~(\ref{Elamb}) that $\lamb$ is
the ratio of the vertical mixing time to the turbulent eddy mixing
time).
\begin{itemize}
	\item  For small $\lamb$, high turbulence $\avk$ and low
	dissipation $\ave$, the dissipation is roughly constant, actually
		\begin{displaymath}
		\ave\approx\frac{1}{\sqrt{\ave^{-2}_{\infty}+15.1/k^3}}\,,
	\end{displaymath}
	as the turbulence decays to~(\ref{Edecay}) on a time scale of
	approximately $2(\eta^2/\ave)^{1/3}$.

\item For large $\lamb$ the vertical mixing time is relatively rapid
and the turbulence decays with a different power law for some time.
We find that $\ave\propto\avk^{2.11}$ which is only a little different
from~(\ref{Edecay}).  The rate of decay towards~(\ref{Edecay}) is
relatively slow,
\begin{displaymath}
	(\avk,\ave)\approx A\left(t^{-0.90},0.90\,t^{-1.90}\right)\,,
\end{displaymath}
and forms a long lasting transient.
\end{itemize}

The above results are for stationary water.
Instead, if the water is moving with uniform velocity on a horizontal
bed then the characteristics of the decaying bulk motion and
turbulence are different in detail.
We seek solutions of the model~(\ref{mdl3}) in the form $\avk\propto
t^{-2}$ and $\ave\propto t^{-3}$, as before, but now with $\avu\propto
t^{-1}$.
Substituting and solving for the constants of proportionality the
model~(\ref{mdl3}) may be rewritten as a generalised eigenvalue
problem for $\lamb$.  It is then straightforward to determine the only
positive solution is
\begin{eqnarray}&&
	\avu \sim 5.52\,\eta t^{-1}\,,\quad
	\avk \sim 20.4\,\eta^2t^{-2}\,,\quad
	\ave \sim 41.1\,\eta^2t^{-3}\,,\quad
	\nonumber\\&&
	\nut \sim 0.91\,\eta^2t^{-1}\,,\quad
	\lamb\sim 0.199\,.
	\label{Eslowdown}
\end{eqnarray}
Numerical solutions show that there are long lasting transients of a
similar nature to those mentioned above for stationary water.  We do
not elaborate further as this class of solutions are less likely to
be of interest in applications.

\subsection{Turbulence in balance with fluid flow}

Water flowing down a slope generates turbulence that provides the drag
to balance the gravitational forcing.
Let the downward bed slope be~$\theta\neq 0$, but as in the previous
subsection assume are no variations in $x$, that is, just a mean flow
($\avu\neq 0$) with no surface waves ($\eta=\mbox{const}$).
Then it is consistent to seek solutions of the model~(\ref{mdl3}) in
the form $\avu\propto \eta^{1/2}\theta^{3/2}$, $\avk\propto
\eta\theta$ and $\ave\propto \eta^{1/2}\theta^{3/2}$.
Substituting and solving for the constants of proportionality leads to
a nonlinearly perturbed eigenvalue problem for $\lamb$ which is solved
iteratively to give
\begin{eqnarray}
	&&
    \avu \approx 3.16\,\eta^{1/2}(g\theta)^{1/2}\,,\quad
	\avk \approx 1.95\,\eta g\theta\,,\quad
	\ave \approx 1.27\,\eta^{1/2}(g\theta)^{3/2}\,,\quad
	\nonumber\\&&
	\nut \approx 0.270\,\eta^{3/2}(g\theta)^{1/2}\,,\quad
	\lamb\approx 0.217\,.
	\label{Eslope}
\end{eqnarray}
We expect this flow to be established on a time scale of the vertical
mixing which is
\begin{displaymath}
	T_{\mbox{\scriptsize mix}}=\frac{\eta^2}{\nut}
	\approx 4\sqrt{\frac{\eta}{g\theta}}\,.
\end{displaymath}

Another interesting balance occurs when we assume that the production
of turbulence parameters, $\avk$ and $\ave$, equals their dissipation
through natural dissipation and bed drag.
This leads to a reduced model in the form of St~Venant's equation used
in open channel flow.
Assume that at all times the production and dissipation of $\avk$ and
$\ave$ in the first lines on the right-hand sides of~(\ref{mdl3k})
and~(\ref{mdl3e}).
That is, assume the bed slope is small enough and the flow is evolving
slowly enough that spatial and temporal gradients are negligible.
Then seek a balance with $\avk\propto\avu^2$ and $\ave\propto\avu^3$
to find
\begin{equation}
	\avk \approx 0.224\,\avu^2\,,\quad
	\ave \approx 0.0453\,\avu^3/\eta\,,\quad
	\nut \approx 0.0992\,\eta\avu\,,\quad
	\lamb\approx 0.184\,.
	\label{Evenant}
\end{equation}
With this balance the momentum equation~(\ref{mdl3u}) becomes
\begin{equation}
	\Dt\avu = -0.111\frac{\avu^2}{\eta}
	+0.94\,g\left( \theta-\Dx\eta \right)
	-1.12\,\avu\Dx\avu
	-0.018\frac{\avu^2}{\eta}\Dx\eta\,,
	\label{Estvenant}
\end{equation}
which has exactly the same form as St Venant's equation for open
channel flow except for the small term $\avu^2\eta_x/\eta$.
The three significant coefficients are worthy of comment: the
self-advection coefficient of $1.12$ accounts for the vertical
nonuniformity of the velocity profile with mean $\avu$; the influence
of gravity is reduced to $0.94\,g$ because, as explained earlier
for~(\ref{mdl2u}), the response of the fluid flow is not constant in
the vertical so some of gravitational forcing is not used; and lastly
the bed drag $\avu^2/\eta$ has coefficient $0.111$ which is larger
than typical values.
However, note that such drag coefficients have to vary depending upon
the roughness of the channel bottom as often expressed by different
roughness coefficients in Manning's law \cite[p246,e.g.]{Bedient88} or
\cite[p137,e.g.]{Kay98}.
We surmise that the flows we describe with model~(\ref{mdl3}) have
strong mixing in the vertical due to strong turbulence generated by a
rough channel bed or other extremely turbulent flows such as breaking
waves or dam spillways.

\subsection{Simulate a dam breaking}
\label{SSdam}

One of the canonical flows of shallow water occurs after a dam
breaks.  Here we simulate such a flow and resolve the water slumping
downstream and becoming extremely turbulent as it does so.  For
simplicity we use the model~(\ref{mdli}) reported in the introduction.

Imagine a dam at $x=0$ initially holding back water of non-dimensional
depth $\eta=1$.
At time $t=0$ the dam breaks and releases the water to rush downstream.
To avoid overly poor conditioning in the numerics we let the water in
front of the dam be of depth $\eta=0.2$ (all quantities will no be
non-dimensional so that in effect $g=1$).
Also to smooth the initial few time steps we actually set $\eta$ to a
tanh profile that smoothly varied between these extremes such that the
water slope was a maximum of $2$ (rather large under the slowly
varying assumption) at the dam.
The water is assumed initially quiescent, $\avu=0$ throughout, and has
a low level of turbulence, somewhat arbitrarily chosen to be
$\avk=0.001$.
Turbulent dissipation is initially set to $\lamb=0.227$ so that the
balance of decaying turbulence~(\ref{Edecay}) holds throughout.

The model~(\ref{mdli}) is simply discretised on a regular grid in
space-time with a time step of $\Delta t=1/10$ and space step of
$\Delta x=1/6$.
The equations are discretised using a six point stencil, 3 wide in
space and 2 in time, and using second order accurate centred
differencing in both space and time.
This leads to a sparse set of nonlinear implicit equations for the
variables at the new time given their values at any given time.
The nonlinear equations are solved by a few Newton's iterations
requiring that the maximum norm of the residuals be less than
$10^{-4}$.
The model~(\ref{mdli}) is purely hyperbolic, so a small amount of
spatial diffusion is incorporated to help stabilise the simulation.
It was physically appealing to incorporate the turbulent diffusion
$(\nut\avu_x)_x$ into the $\avu$~equation and similarly for for $\avk$
and $\ave$.
Such diffusion made hardly any difference to the simulations as a
whole yet usefully avoided the generation of unphysical and ruinous
spikes in the numerical solution.
The domain of simulation extended from six dam heights to the left
behind the dam to six dam heights downstream.
We integrated for over a time $t=7$ which is long enough for the
disturbance to reach the ends of the computational domain (linear
waves on the dammed water having speed~1).

\begin{figure}[tbp]
	\centering
	\includegraphics{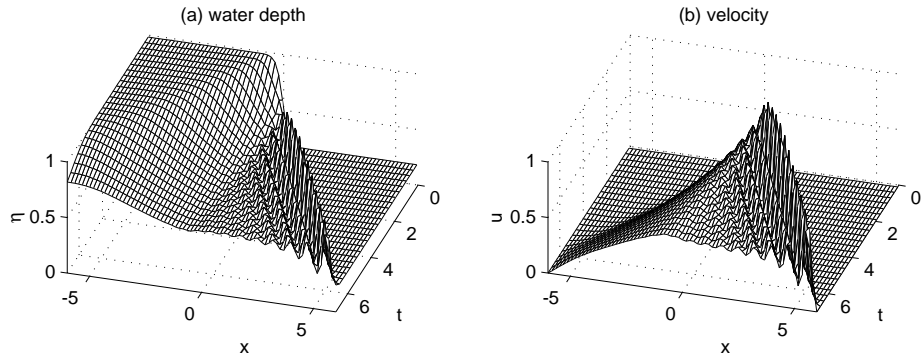}
	\caption{simulation of dam breaking showing (a) the water depth
	$\eta$ and (b) the mean downstream velocity $\avu$.  Observe the
	formation of a bore with superposed waves.}
	\label{fig:damhu}
\end{figure}
\begin{figure}[tbp]
	\centering
	\includegraphics{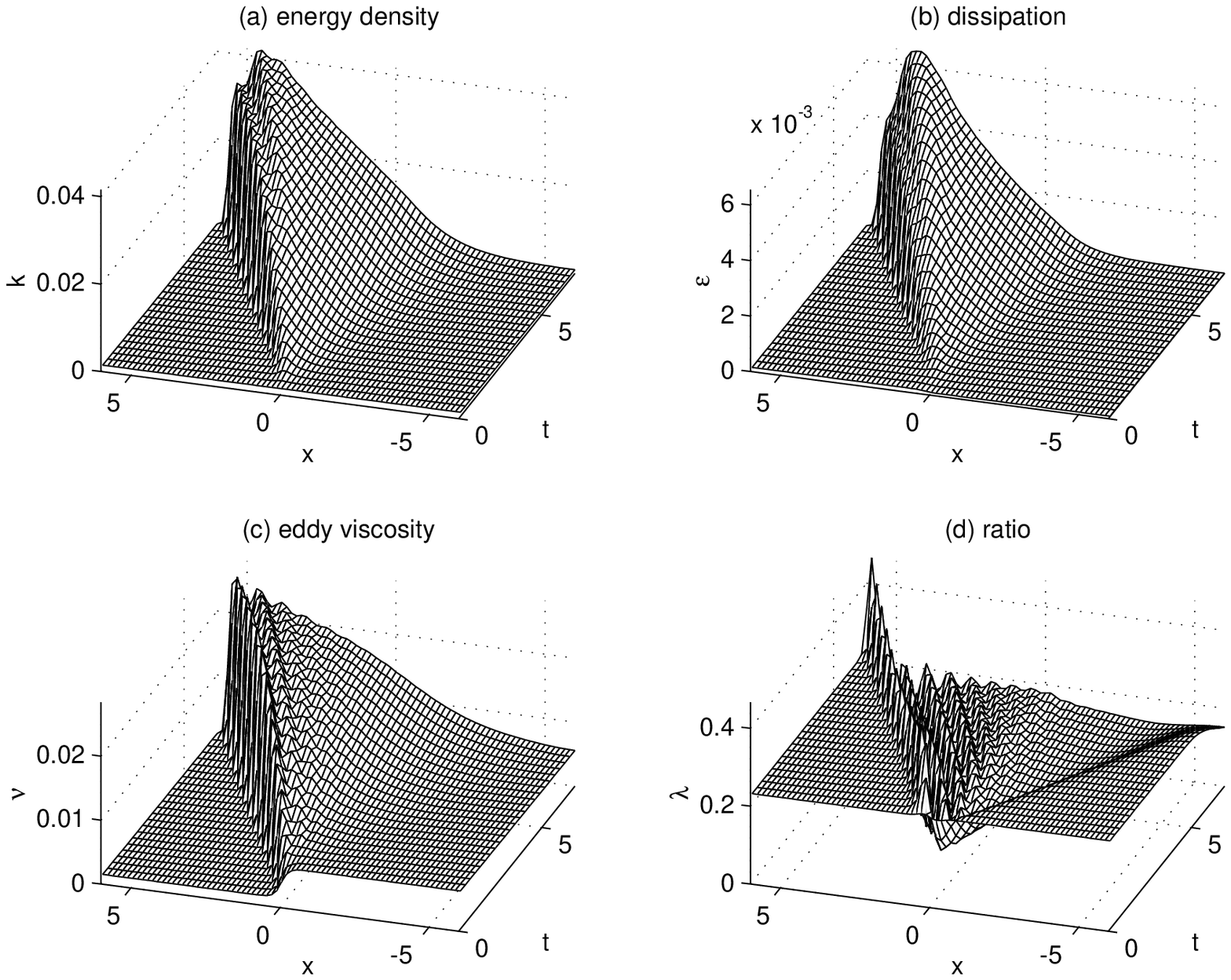}
\caption{simulation of dam breaking showing the time evolution of the
turbulence parameters (note the view point is rotated from
Figure~\ref{fig:damhu}): (a) the turbulent energy density $\avk$ is
highest just a little behind the bore and then tails away; (b) the
turbulent dissipation $\ave$ behaves similarly; (c) the turbulent eddy
viscosity $\nut$ is greatest at the front of the bore; and (d) the
parameter $\lamb$, apart from a peak at he front of the bore, is
generally depressed from the decaying balance value of $0.227$.}
	\label{fig:damturb}
\end{figure}

The results of this simulation are shown in Figures~\ref{fig:damhu}
and~\ref{fig:damturb}.
Observe that when the dam breaks, the water slumps down and rushes
downstream in a turbulent bore.
The bore appears undular but may be evolving towards a series of
solitary waves---they cannot be differentiated on this time-scale.
The turbulent structure shown in Figure~\ref{fig:damturb} has
interesting features.
Apparently the energy density peaks a little behind the bore, then
decays approximately linearly with distance.
It appears that up to time $t=7$ the generation of turbulence is still
significantly greater than its decay as the peak is still growing.
The turbulent dissipation and eddy viscosity behave similarly, though
the eddy viscosity appears to peak much closer to the front of the
turbulent bore.
The parameter $\lamb$ plotted in Figure~\ref{fig:damturb}(d) reaffirms
that relatively small values appear relevant to flows of interest.
The peak of $\lamb$ at the front of the bore predicts that there is a
lot of vertical mixing at the front, but less so behind the bore where
$\lamb$ is smaller.
All of the above seem physically reasonable.

This model that we have derived and solved in some cases of interest
explicitly accounts for the spatio-temporal variations of the
intensity and broad nature of the turbulence underlying the flow of
shallow water.

\paragraph{Acknowledgement}
This research was supported by a grant from the University of Southern
Queensland, by the Australian Research Council, and by the Volkswagen
Foundation, Germany, I/69449.

\addcontentsline{toc}{section}{References}
\bibliographystyle{plain}
\bibliography{new,ajr,bib}

\begin{thebibliography}{10}

\bibitem{Abramowitz64}
M.~Abramowitz and I.A. Stegun, editors.
\newblock {\em Handbook of mathematical functions}.
\newblock Dover, 1965.

\bibitem{AnNo82}
R.~Arnold and J.~Noye.
\newblock Numerical modelling of long waves.
\newblock In J.~Noye, editor, {\em Numerical Solutions of Partial Differential
  Equations}, pages 437--453. North-Holland, 1982.

\bibitem{AnNo84}
R.~Arnold and J.~Noye.
\newblock On the performance of turbulent energy closure schemes of wind driven
  flows in shallow seas.
\newblock In J.~Noye, editor, {\em Computational Techniques and Applications:
  CTAC-83}, pages 425--437. North-Holland, 1984.

\bibitem{AnNo86}
R.~Arnold and J.~Noye.
\newblock Open boundary conditions for a tidal and storm surge model of bass
  strait.
\newblock In J.~Noye, editor, {\em Computational Techniques and Applications:
  CTAC-85}, pages 503--518. North-Holland, 1986.

\bibitem{Balakotaiah92}
V.~Balakotaiah and H.C. Chang.
\newblock Dispersion of chemical solutes in chromatographs and reactors.
\newblock {\em Phil Trans R Soc Lond A}, 351:39--75, 1995.

\bibitem{Bedient88}
P.B. Bedient and W.C. Huber.
\newblock {\em Hydrology and floodplain analysis}.
\newblock Addison-Wesley, 1988.

\bibitem{Bertschy83}
J.R. Bertschy, R.W. Chin, and F.H. Abernathy.
\newblock High-strain-rate free-surface boundary-layer flows.
\newblock {\em J. Fluid Mech.}, 126:443--461, 1983.

\bibitem{Birkhoff69}
G.~Birkhoff and G.-C. Rota.
\newblock {\em Ordinary Differential Equations}.
\newblock Xerox College Publishing, 1969.

\bibitem{Carr81}
J.~Carr.
\newblock {\em Applications of centre manifold theory}, volume~35 of {\em
  Applied Math Sci}.
\newblock Springer-Verlag, 1981.

\bibitem{Coullet83}
P.H. Coullet and E.A. Spiegel.
\newblock Amplitude equations for systems with competing instabilities.
\newblock {\em SIAM J. Appl. Math.}, 43:776--821, 1983.

\bibitem{Durbin91}
P.A. Durbin.
\newblock Near-wall turbulence closure modeling without ``damping functions''.
\newblock {\em Theoret. Comput. Fluid Dynamics}, 3:1--13, 1991.

\bibitem{Durbin93}
P.A. Durbin.
\newblock Application of a near-wall turbulence model to boundary layers and
  heat transfer.
\newblock {\em Int. J. Heat and Fluid Flow}, 14(4):316--323, 1993.

\bibitem{Durbin93b}
P.A. Durbin.
\newblock A reynolds stress model for near-wall turbulence.
\newblock {\em J. Fluid Mech.}, 249:465--498, 1993.

\bibitem{Fredsoe92}
J.~Fredsoe and R.~Deigaard.
\newblock {\em Mechanics of coastal sediment transport}, volume~3 of {\em Adv.
  Series on Ocean Eng.}
\newblock World Sci., 1992.

\bibitem{Gallay93}
Th. Gallay.
\newblock A center-stable manifold theorem for differential equations in banach
  spaces.
\newblock {\em Commun. Math. Phys}, 152:249--268, 1993.

\bibitem{Gibson89}
M.M. Gibson and W.~Rodi.
\newblock {Simulation of free surface effects on turbulence with a Reynolds
  stress model}.
\newblock {\em J.~Hydraulic Res.}, 27:233--244, 1989.

\bibitem{Haken83}
H.~Haken.
\newblock {\em Synergetics, An introduction}.
\newblock Springer, Berlin, 1983.

\bibitem{Hanjalic72}
K.~Hanjali\'c and B.E. Launder.
\newblock A reynolds stress model of turbulence and its applications to thin
  shear flows.
\newblock {\em J. Fluid Mech.}, 52:609--638, 1972.

\bibitem{Hartman82}
P.~Hartman.
\newblock {\em Ordinary Differential Equations (2nd Edition)}.
\newblock Birkh\"auser, 1982.

\bibitem{Hinch91}
E.J. Hinch.
\newblock {\em Perturbation methods}.
\newblock Cambridge texts in Applied Mathematics. CUP, 1991.

\bibitem{Hodges99}
B.R. Hodges and R.L. Street.
\newblock On simulation of turbulent nonlinear free-surface flows.
\newblock {\em J. Computational Physics}, 151:425---457, 1999.

\bibitem{Kay98}
M.~Kay.
\newblock {\em Practical hydraulics}.
\newblock E. \& F.N. Spon, 1998.

\bibitem{Keller88}
R.J. Keller and W.~Rodi.
\newblock Prediction of flow characteristics in main channel/flood plain flows.
\newblock {\em J Hydraulic Res}, 26(4):425--442, 1988.

\bibitem{Krogstad94}
P.-A. Krogstad and R.A. Antonia.
\newblock Structure of turbulent boundary layers on smooth and rough walls.
\newblock {\em J. Fluid Mech.}, 277:1--21, 1994.

\bibitem{Launder75}
B.E. Launder, G.J. Reece, and W.~Rodi.
\newblock Progress in the development of a reynolds-stress turbulence closure.
\newblock {\em J. Fluid Mech.}, 68:537--566, 1975.

\bibitem{Mei89}
C.C. Mei.
\newblock {\em The applied dynamics of ocean surface waves}, volume~1 of {\em
  Advanced series on ocean engineering}.
\newblock World Scientific, 1989.
\newblock 2nd printing.

\bibitem{Mei94}
Z.~Mei and A.J. Roberts.
\newblock Equations for turbulent flood waves.
\newblock In A.~Mielke and K.~Kirchg\"assner, editors, {\em Structure and
  dynamics of nonlinear waves in fluids}, pages 342--352. World Sci, 1995.

\bibitem{Mercer90}
G.N. Mercer and A.J. Roberts.
\newblock A centre manifold description of contaminant dispersion in channels
  with varying flow properties.
\newblock {\em SIAM J. Appl. Math.}, 50:1547--1565, 1990.

\bibitem{Mercer94a}
G.N. Mercer and A.J. Roberts.
\newblock A complete model of shear dispersion in pipes.
\newblock {\em Jap. J. Indust. Appl. Math.}, 11:499--521, 1994.

\bibitem{Mohammadi94}
B.~Mohammadi and O.~Pironneau.
\newblock Analysis of the $k$-$epsilon$ turbulence model.
\newblock 1994.

\bibitem{Peregrine72}
D.H. Peregrine.
\newblock Equations for water waves and the approximations behind them.
\newblock In R.E. Meyer, editor, {\em Waves On Beaches And Resulting Sediment},
  pages 95--121. Academic Press, 1972.

\bibitem{Prokopiou91b}
Th. Prokopiou, M.~Cheng, and H.C. Chang.
\newblock {Long waves on inclined films at high Reynolds number}.
\newblock {\em J. Fluid Mech.}, 222:665--691, 1991.

\bibitem{Rastogi78}
A.K. Rastogi and W.~Rodi.
\newblock Prediction of heat and mass transfer in open channels.
\newblock {\em J Hydraulics Div}, 104:397--419, 1978.

\bibitem{Roberts85b}
A.J. Roberts.
\newblock Simple examples of the derivation of amplitude equations for systems
  of equations possessing bifurcations.
\newblock {\em J.~Austral. Math. Soc. B}, 27:48--65, 1985.

\bibitem{Roberts88a}
A.J. Roberts.
\newblock The application of centre manifold theory to the evolution of systems
  which vary slowly in space.
\newblock {\em J. Austral. Math. Soc. B}, 29:480--500, 1988.

\bibitem{Roberts89b}
A.J. Roberts.
\newblock Appropriate initial conditions for asymptotic descriptions of the
  long term evolution of dynamical systems.
\newblock {\em J. Austral. Math. Soc. B}, 31:48--75, 1989.

\bibitem{Roberts92c}
A.J. Roberts.
\newblock Boundary conditions for approximate differential equations.
\newblock {\em J. Austral. Math. Soc. B}, 34:54--80, 1992.

\bibitem{Roberts93}
A.J. Roberts.
\newblock The invariant manifold of beam deformations. part 1:the simple
  circular rod.
\newblock {\em J.~Elas.}, 30:1--54, 1993.

\bibitem{Roberts94c}
A.J. Roberts.
\newblock Low-dimensional models of thin film fluid dynamics.
\newblock {\em Phys.\ Letts.~A}, 212:63--72, 1996.

\bibitem{Roberts97a}
A.J. Roberts.
\newblock Low-dimensional modelling of dynamical systems.
\newblock Technical report, [\url{http://xxx.lanl.gov/abs/chao-dyn/9705010}],
  1997.

\bibitem{Roberts96a}
A.J. Roberts.
\newblock Low-dimensional modelling of dynamics via computer algebra.
\newblock {\em Comput. Phys. Comm.}, 100:215--230, 1997.

\bibitem{Roberts96b}
A.J. Roberts.
\newblock An accurate model of thin 2d fluid flows with inertia on curved
  surfaces.
\newblock In P.A. Tyvand, editor, {\em Free-surface flows with viscosity},
  volume~16 of {\em Advances in Fluid Mechanics Series}, chapter~3, pages
  69--88. Comput Mech Pub, 1998.

\bibitem{Roberts97b}
A.J. Roberts.
\newblock Computer algebra derives correct initial conditions for
  low-dimensional dynamical models.
\newblock Submitted to Comput. Phys. Comm., February 1999.

\bibitem{Rodi80}
W.~Rodi.
\newblock {\em Turbulence Models and Their Applications in Hydraulics}.
\newblock International Association of Hydraulic Research, 1980.

\bibitem{Shiono91}
K.~Shiono and D.W. Knight.
\newblock Turbulent open-channel flows with variable depth across the channel.
\newblock {\em J. Fluid Mech.}, 222:617--646, 1991.

\bibitem{Speziale91}
C.G. Speziale.
\newblock Analytical methods for the development of reynolds-stress closures in
  turbulence.
\newblock {\em Annu Rev Fluid Mech}, 23:107--157, 1991.

\bibitem{Watt94b}
S.D. Watt and A.J. Roberts.
\newblock The accurate dynamic modelling of contaminant dispersion in channels.
\newblock {\em SIAM J Appl Math}, 55(4):1016--1038, 1995.

\end{thebibliography}

\appendix
\section{Comments on theory in this application}
\label{Scom}

This appendix addresses the connection to the rigorous theory of centre
manifolds in this application.

It must be emphasised that throughout this paper we describe the
application of centre manifold techniques and \emph{not} the
application of the rigorous centre manifold theory.
There are two main reasons for this which we elaborate on below.

Firstly, here we construct an infinite dimensional centre manifold.
At each point in $x$ there are four degrees of freedom, parametrised
by $\eta$, $\bar u$, $\bar k$ and $\bar\epsilon$; but there are an
infinitude of $x$ positions and so there is an infinite number of
degrees of freedom in the mathematical model.  However, there is
currently very little theory on infinite dimensional centre manifolds
appearing via slowly-varying approximations \cite[e.g.]{Gallay93}, and
what there is does not rigorously apply here, nor does it apply to
many other physically interesting models such as dispersion in pipes
\cite{Mercer94a}, laminar long-wave, thin-film flows
\cite{Roberts94c}, and the dynamics in flow reactors
\cite{Balakotaiah92}.  Instead we use the formal techniques of
constructing complete low-dimensional models
\cite{Coullet83,Roberts85b,Roberts88a,Roberts89b,Roberts92c},
techniques suggested and developed by standard applications of the
theory.  We expect that eventually theory will be developed which
supports the application of centre manifold concepts to slowly-varying
approximations.

But there is a second obstacle to supporting this model with theory.
In standard applications of centre manifold theory the nonlinear terms
in the original problem are required to be smooth in the neighbourhood
of the equilibrium under consideration (here the origin, a state of no
flow and no turbulence).  However, here many nonlinear terms are
definitely not smooth; for example, turbulent dispersion terms such as
$\pard{~}{x}\left( \Cm\frac{k^2}\epsilon \pard{u}{x}\right)$ and
interaction terms such as $\Cet{\epsilon^2\over k}$ are unbounded as
$(u,k,\epsilon)\to\vec 0$.  In rigorous applications of centre
manifold theory, one may choose the various critical modes and
parameters to have any set of relative orders of magnitude.  The
resulting asymptotic expressions are the same \cite{Roberts85b}, it is
only the sequence in which the terms appear that changes with a change
in relative orders of magnitude.  Indeed, this reflects a very
desirable property of a modelling procedure, namely that the results
are essentially independent of arbitrary human-made assumptions (such
as order of magnitude) in the analysis.  However, in this application
the highly nonsmooth nature of the original equations means that in
order to apply the centre manifold techniques we need to choose
carefully the various orders of magnitudes of the variables and
parameters, via~(\ref{deltadefn}).  The aim, as in all asymptotic
analyses, is to obtain a tractable and physically relevant leading
order problem.  The techniques of centre manifold theory are then
applied, it is just that the current centre manifold theorems cannot
give \emph{rigorous} justification.

Notwithstanding these theoretical limitations we consider the
systematic techniques are applicable because of the attractivity of
the manifold of equilibria, $\cM_0$, of the linear operator $\cL$.
In the spirit of centre manifold theory, we claim that the ``small
nonlinear'' terms on the right-hand side of~(\ref{lneqn}) just perturb
the shape of $\cM_0$ to a nearby manifold $\cM$ and perturb the
evolution thereon.
Thus our last task, and the one fulfilled in this Appendix, is to
prove the exponential decay to $\cM_0$.

Linearising the {\keps} equations~(\ref{lneqn}) about $\cM_0$
\begin{equation}
    (p_0,\u_0)
  = \left(g(h-y),\cubey U(x,t),0,H(x,t),\cubey K(x,t),E(x,t)\right)\,,
        \label{EqMan}
\end{equation}
we obtain $(0,\partial\u/\partial t)=\cL(p,\u)$ where the linear
operator $\cL$ is
\begin{displaymath}
\left[\matrix{0 &0 &\pard{\cdot}{y} & 0 & 0 &0\cr 0
&\pard{~}{y}\left(\nu_0\pard{\cdot}{y}\right) & 0 & 0
&\Cm\pard{~}{y}\left(\pard{u_0}{y}{2k_0\over\epsilon_0}\cdot\right)
&-\Cm\pard{~}{y}\left(\pard{u_0}{y}{k_0^2\over\epsilon_0^2}\cdot\right)\cr
-\pard{\cdot}{y} &0&2\pard{~}{y}\left(\nu_0\pard{\cdot}{y}\right) & 0
&-\frac23\pard{\cdot}{y}
&0\cr
0 & 0&0&0&0&0\cr 0 &
0&0&0
&{\Cm\over\sk}\pard{~}{y}\left({k_0^2\over\epsilon_0}\pard{\cdot}{y}+
{2k_0\over\epsilon_0}\pard{k_0}{y}\cdot\right)
&-\frac{\Cm}{\sk}\pard{~}{y}
\left(\pard{k_0}{y}{k_0^2\over\epsilon_0^2}\cdot\right)\cr
0 & 0&0&0&0
&{\Cm\over\se}\pard{~}{y}
\left({k_0^2\over\epsilon_0}\pard{\cdot}{y}\right)
}\right],
\end{displaymath}
subject to boundary conditions
\begin{eqnarray}
        p+\frac{2}{3}k & = & 0\quad\mbox{on $y=h$\,,}
        \label{BCpa} \\
        u & = & 0\quad\mbox{on $y=0$\,,}
        \label{BCu0} \\
        \frac{\partial u}{\partial y}-\frac{u}{3h} & = & 0
		\quad\mbox{on $y=h$\,,}
        \label{BCuh} \\
        v & = & 0\quad\mbox{on $y=0$\,,}
        \label{BCv} \\
        k & = & 0\quad\mbox{on $y=0$\,,}
        \label{BCk0} \\
        \frac{\partial k}{\partial y}-\frac{k}{3h} & = & 0
		\quad\mbox{on $y=h$\,,}
        \label{BCkh} \\
        y^{2/3}\frac{\partial\epsilon}{\partial y} & \to & 0\quad\mbox{as
        $y\to0$\,,}
        \label{BCe0} \\
        \frac{\partial\epsilon}{\partial y} & = & 0\quad\mbox{on
        $y=h$\,.}
        \label{BCeh}
\end{eqnarray}

We seek solutions proportional to $\exp(\lambda t)$.
The first thing to note is that we address a generalised eigen-problem
\[ \cL\left[
\begin{array}{c}
        p_0  \\
        \u
\end{array}
\right]=\left[
\begin{array}{c}
        0  \\
        \lambda \u
\end{array}
\right]
\]
as the first row of $\cL$ comes from the continuity equation.

Thus the first row, with the boundary condition~(\ref{BCv}), gives
$v=0$ for any eigenvalue $\lambda$.
Furthermore, the third row, from the vertical momentum equation with
the pressure boundary condition~(\ref{BCpa}), then gives that
$p=-\frac{2}{3}k$ for any $\lambda$.
These considerations give no constraint on the eigenvalue $\lambda$.
Ignoring the $p$ and $v$ components of the eigen-problem then lead to
a standard eigen-problem; one where the operator is in block,
upper-triangular form.
Thus consider the components in turn, starting from the last, and we
show that all eigenvalues must be non-positive and thus $\cM_0$ is
attractive.
\begin{itemize}

 \item For any eigen-function $\u$, if turbulent dissipation
$\epsilon\neq0$, then
\begin{eqnarray}
  \frac{h^{4/3}}{T\se}
  \pard{~}{y}\left(y^\ftt\pard{\epsilon }{y}
  \right)&=&\lambda \epsilon \,,
  \label{EP2}
\end{eqnarray}
where
\[ T=\frac{9h^2}{16\nut}\,,
\]
is the time-scale of cross-depth turbulent diffusion.
Multiplying~(\ref{EP2}) by $\epsilon$, $\int_0^h\cdots dy$, and
integrating by parts we deduce
\[ \lambda=-\frac{h^{4/3}}{T\se}\frac
{\int_0^hy^{2/3}\left(\pard\epsilon y\right)^2dy}
{\int_0^h\epsilon^2dy}
\leq 0\,,
\]
provided that the boundary conditions (\ref{BCeh})  (assuming
$\epsilon\not\to \infty$ for $y\to h$) and
\begin{equation}
        \epsilon\frac{\partial\epsilon}{\partial y}=o\left(y^{-2/3}\right)
        \quad\mbox{as $y\to0$,}
        \label{BCe0x}
\end{equation}
are satisfied. The equality $\lambda=0$ holds if and only if
$\frac{\partial\epsilon}{\partial y}\equiv0$, which leads
to
$$
     \epsilon=E(x,t)\,,
$$
a function independent of $y$. Since equation~(\ref{EP2}) indicates
nontrivial solutions near $y=0$ are of the form $\epsilon\sim A+By^{1/3}$,
then the condition~(\ref{BCe0x}) is effectively equivalent to~(\ref{BCe0}).

Further, apply Sturm-Liouville theory to~(\ref{EP2}) under
boundary conditions~(\ref{BCeh}) and (\ref{BCe0}).
Changing the vertical variable from $y=hz^3$ to $z$,
the eigenvalue problem becomes
\[
     \pard{^2\epsilon}{z^2} = 9T\se\lambda z^2 \epsilon\,,
\]
(a form of Bessel's equation~\cite[Eq.9.1.51]{Abramowitz64}) with the
following normal separate boundary conditions
\[
        \frac{\partial\epsilon}{\partial z}= 0 \quad\mbox{on $y=0$ and
        $y=h$.}
\]
Applying standard Sturm-Liouville theory, see for example Birkhoff \&
Rota \cite[pp.~296]{Birkhoff69} or Hartman
\cite[pp.~337ff]{Hartman82}, we see that the eigenvalues are discrete
and must tend to infinity,
\[
   0=\lambda_1>\lambda_2>\cdots \to -\infty\,.
\]
Thus, linearly, solutions in the neighbourhood of the manifold $\cM_0$
are attracted exponentially quickly to it (at a rate at least as fast as
$\exp(\lambda_2 t)$).

Sturm-Liouville theory may be also applied directly for the $u$ and
$k$ components to show that any eigenvalues associated primarily with them
are discrete.  We do not record the details in the following.

 \item Similarly, for any eigen-function $\u$, if turbulent dissipation
$\epsilon=0$ but the turbulent energy $k\neq0$, then
\begin{eqnarray}
  \frac{h^{4/3}}{T\sk}
  \pard{~}{y}\left(y^\ftt\pard{k}{y}
  +\frac{2k}{3y^{1/3}} \right)&=&\lambda k \,.
  \label{EP3}
\end{eqnarray}
Multiplying this by $y^{2/3}$ we rewrite it as
\[   \frac{h^{4/3}}{T\sk}
y^{-1/3}\pard{~}y\left[y^2\pard{~}y\left(y^{-1/3}k\right)\right]
=\lambda y^{2/3}k\,.
\]
Multiplying by $k$, $\int_0^h\cdots dy$ and integrating by parts we deduce
\[ \lambda=-\frac{h^{4/3}}{T\sk}\frac
{\int_0^hy^{2}\left[\pard{~} y\left(y^{-1/3}k\right)\right]^2dy}
{\int_0^h y^{2/3}k^2dy}
\leq 0\,,
\]
provided (\ref{BCkh}) and
\begin{equation}
        k=o\left(y^{1/12}\right)\quad\mbox{as $y\to0$\,,}
        \label{BCk0x}
\end{equation}
are satisfied. The equality $\lambda=0$ holds here if and only if
$\pard{~} y\left(y^{-1/3}k\right)\equiv0$, which together with
boundary condition (\ref{BCk0}) implies
$$
     k=K(x,t)\cubey
$$
with a function $K(x,t)$ independent of $y$. Since the indicial equation
of~(\ref{EP3}) indicates nontrivial solutions near $y=0$ are of the form
$k\sim Ay^{-2/3}+By^{1/3}$, then the boundary condition~(\ref{BCk0x}) is
equivalent to~(\ref{BCk0}).

\item The only possible eigenvalue associated with non-zero $\eta$
is $0$.

\item Lastly, for any eigen-function $\u$, if
$\epsilon=k=\eta=0$ but the horizontal velocity $u\neq0$, then
\begin{eqnarray}
  \frac{h^{4/3}}{T}
  \pard{~}{y}\left(y^\ftt\pard{u}{y} \right)&=&\lambda u \,,
  \label{EP4}
\end{eqnarray}
and we rewrite this as
\[   \frac{h^{4/3}}{T}
y^{-1/3}\pard{~}y\left[y^{4/3}\pard{~}y\left(y^{-1/3}u\right)\right]
=\lambda u\,.
\]
Multiplying by $u$, $\int_0^h\cdots dy$ and integrating by parts we deduce
\[ \lambda=-\frac{h^{4/3}}{T}\frac
{\int_0^hy^{4/3}\left[\pard{~} y\left(y^{-1/3}u\right)\right]^2dy}
{\int_0^h u^2dy}
\leq 0\,,
\]
provided (\ref{BCuh}) and
\begin{equation}
        u=o\left(y^{1/6}\right)\quad\mbox{as $y\to0$\,,}
        \label{BCu0x}
\end{equation}
are satisfied.  Similarly to the $\epsilon$ and $k$ equations, the
equality $\lambda=0$ holds here if and only if $\pard{~}
y\left(y^{-1/3}u\right)\equiv0$.  This and the boundary condition
(\ref{BCk0}) yields a unique solution
\[
     u=U(x,t)\cubey
\]
with $U(x,t)$ independent of $y$. Since the indicial equation
of~(\ref{EP4}) indicates nontrivial solutions near $y=0$ are of the form
$u\sim A+By^{1/3}$, then the boundary condition~(\ref{BCu0x}) is
effectively equivalent to~(\ref{BCu0}).

\end{itemize}

We have proven: \emph{if the boundary conditions
(\ref{BCpa}--\ref{BCeh}) are satisfied, then, except for the four-fold
eigenvalue zero whose eigenfunctions span $\cM_0$, the eigenvalues of
$\cL$ are negative and bounded away from $0$.  } Thus we expect the
manifold~(\ref{EqMan}) is locally attractive.  Further, the time-scale
of this attraction is the cross-depth turbulent diffusion time-scale
$T$.

\section{Computer algebra}
\label{Salg}

Here we list the \textsc{reduce}\footnote{At the time of writing,
information about \textsc{reduce} was available from Anthony C.\
Hearn, RAND, Santa Monica, CA 90407-2138, USA.
\url{mailto:reduce@rand.org}} computer algebra program used to derive
the long-wave models of turbulent flow.

The algorithm is the iterative algorithm described in
\cite{Roberts96a,Roberts97b} adapted to this difficult asymptotic
problem.
The program refines the description of the centre manifold and the
evolution thereon until the residual of the governing differential
equations are driven to zero, to some asymptotic error.
The key to the correctness of the results is then in the correct
coding of the residuals---see inside the iterative loop.

Note that because the thickness of the film is continuously varying in
space and time, and because of the cube-root structure in the
vertical, it is convenient to work with equations in terms of a scaled
vertical coordinate $z=\sqrt[3]{y/\eta}$ so that the free surface of
the film is always $z=1$.
However, the turbulence equations are not explicitly rewritten in this
new coordinate because the computer handles all the necessary details
of the transformation.

\verbatimlisting{cbrt.red}

\end{document}